\documentclass[conference]{IEEEtran}
\ifCLASSOPTIONcompsoc
  \usepackage[nocompress]{cite}
\else
  \usepackage{cite}
\fi
\ifCLASSINFOpdf
\else
\fi
\usepackage{amsmath}
\usepackage{amssymb}
\usepackage{booktabs}
\usepackage{caption}
\usepackage{multirow}
\usepackage{makecell}
\usepackage{placeins}
\usepackage{graphicx}
\usepackage{pgf-pie} % For pie charts
\usepackage{caption} % For subcaptions
\usepackage{subcaption}
\usepackage{tikz} % For custom placements
\usetikzlibrary{positioning}
\usepackage{xtab}
\usepackage{pgfplots}
\pgfplotsset{compat=1.17}
\usepackage{pgfplotstable}
\usepackage[linesnumbered,ruled,vlined]{algorithm2e}
\usepackage{amsmath}
\usepackage{url}
\usetikzlibrary{matrix}
\usepackage[T1]{fontenc}
\usepackage[utf8]{inputenc}

\definecolor{1}{HTML}{003366} 
\definecolor{2}{HTML}{336699} 
\definecolor{3}{HTML}{6699CC} 
\definecolor{4}{HTML}{99CCFF}

\definecolor{blue1}{HTML}{03131f} 
\definecolor{blue2}{HTML}{153f65} 
\definecolor{blue3}{HTML}{70bdf2} 
\definecolor{blue4}{HTML}{c1e3ff}
\definecolor{blue5}{HTML}{edf4fe} 

\definecolor{red1}{HTML}{FF4500} 
\definecolor{red2}{HTML}{FF6347} 
\definecolor{red3}{HTML}{FF8C00} 

\definecolor{green1}{HTML}{2B5A41} 
\definecolor{green2}{HTML}{4A7A5E} 
\definecolor{green3}{HTML}{7B9F80} 
\definecolor{green4}{HTML}{D2C7A3} 
\definecolor{green5}{HTML}{d2f0d5} 
\definecolor{green6}{HTML}{81be4d}

\usepackage{tcolorbox}

\newcounter{insightcounter}
\newcommand{\insight}[1]{
    \stepcounter{insightcounter}
    \begin{tcolorbox}[colback=gray!5!white, colframe=gray!80!black, boxrule=0.3pt, rounded corners, fonttitle=\bfseries]
    \textit{\textbf{Insight \arabic{insightcounter}}}: #1 
    \end{tcolorbox}
}

\newcounter{observationcounter}

\usepackage{algorithmic}
\usepackage{array}
\begin{document}
\title{SoK: Stablecoin Designs, Risks, \\and the Stablecoin LEGO}

\author{
\IEEEauthorblockN{
    Shengchen Ling\IEEEauthorrefmark{1}, 
    Yuefeng Du\IEEEauthorrefmark{1}, 
    Yajin Zhou\IEEEauthorrefmark{2}, 
    Lei Wu\IEEEauthorrefmark{2},
    Cong Wang\IEEEauthorrefmark{1},
    Xiaohua Jia\IEEEauthorrefmark{1},
    Houmin Yan\IEEEauthorrefmark{1} 
}
\IEEEauthorblockA{
    \IEEEauthorrefmark{1}City University of Hong Kong,
    \IEEEauthorrefmark{2}Zhejiang University
}
}
	

%Telephone: (800) 555--1212, Fax: (888) 555--1212}


\IEEEoverridecommandlockouts
\makeatletter\def\@IEEEpubidpullup{6.5\baselineskip}\makeatother
% \IEEEpubid{\parbox{\columnwidth}{
% 		Network and Distributed System Security (NDSS) Symposium 2026\\
% 		24-28 February 2025, San Diego, CA, USA\\
% 		ISBN 979-8-9894372-8-3\\
% 		https://dx.doi.org/10.14722/ndss.2025.[23|24]xxxx\\
% 		www.ndss-symposium.org
% }
% \hspace{\columnsep}\makebox[\columnwidth]{}}

\maketitle

\begin{abstract}

Stablecoins have become significant assets in modern finance, with a market capitalization exceeding USD 246 billion (May 2025). Yet, despite their systemic importance, a comprehensive and risk-oriented understanding of crucial aspects like their design trade-offs, security dynamics, and interdependent failure pathways often remains underdeveloped. This SoK confronts this gap through a large-scale analysis of 157 research studies, 95 active stablecoins, and 44 major security incidents.

Our analysis establishes four pivotal insights: 1) stability is best understood not an inherent property but an emergent, fragile state reliant on the interplay between market confidence and continuous liquidity; 2) stablecoin designs demonstrate trade-offs in risk specialization instead of mitigation; 3) the widespread integration of yield mechanisms imposes a ``dual mandate'' that creates a systemic tension between the core mission of stability and the high-risk financial engineering required for competitive returns; and 4) major security incidents act as acute ``evolutionary pressures'', forging resilience by stress-testing designs and aggressively redefining the security frontier. We introduce the Stablecoin LEGO framework, a quantitative methodology mapping historical failures to current designs. Its application reveals that a lower assessed risk strongly correlates with integrating lessons from past incidents. We hope this provides a systematic foundation for building, evaluating, and regulating more resilient stablecoins.

\end{abstract}


% \fi
\IEEEpeerreviewmaketitle

\section{Introduction}

Digital assets, particularly cryptocurrencies, offer a level of transactional convenience that can surpass traditional systems. However, the pronounced volatility of prominent cryptocurrencies like Bitcoin renders them unsuitable as stable mediums of exchange. This limitation underscores the critical need for stablecoins, which aim to facilitate seamless everyday transactions by maintaining a stable value, thereby providing a reliable store of value amidst market fluctuations and economic turbulence.

Blockchain-based stablecoins have rapidly achieved a market capitalization exceeding USD 246 billion, profoundly influencing both the decentralized finance (DeFi) ecosystem and its intersections with traditional financial systems. Yet, despite this systemic importance, the escalating frequency of security incidents (most notably the Terra event, which caused losses near USD 40 billion~\cite{BRIOLA2023103358}) underscores an urgent challenge. These developments mandate a rigorous, comprehensive understanding of stablecoin design architectures and inherent risk profiles to inform safer practices and guide future innovation.

While prior research has systematically surveyed the broader DeFi landscape~\cite{10.1145/3558535.3559780}, encompassing decentralized exchanges (DEXs)~\cite{10.1145/3570639}, yield aggregators~\cite{9805523}, governance~\cite{10.1145/3558535.3559794}, and security incidents~\cite{10179435}, and while specific studies have addressed stablecoins~\cite{ito2020stablecoin,10.1145/3419614.3423261,Klages2021In,10634419}, these analyses often lack contemporary advancements or are confined by primarily economic viewpoints. Consequently, a significant lacuna persists: the absence of an integrated, interdisciplinary framework for systematically understanding stablecoin design, quantifying associated risks, and evaluating their ecosystem-wide implications. 

\noindent\textbf{Our work.} This SoK confronts this lacuna directly. Grounded in a large-scale analysis of 157 research studies, 95 operational stablecoins, and 44 major security incidents, we deliver a holistic systematization of the stablecoin ecosystem. Our work is built upon four pivotal insights that challenge prevailing assumptions and provide a new lens for understanding stablecoin security.

Our analysis begins by establishing a foundational premise: \textbf{for a stablecoin, stability is an emergent and fragile state, not an inherent property.} Distinct from other DeFi tokens, a stablecoin’s sole mission is peg stability. Our analysis reveals this is not a static feature but an adaptive socio-technical process. It relies fundamentally on two market-validated conditions: sustained market confidence, earned through transparent collateral and robust mechanisms, and effective convertibility (liquidity) into its reference value.

This inherent fragility forces designers into a landscape of difficult trade-offs, where we find that \textbf{design choices result in risk specialization rather than complete risk elimination.} Stablecoins typically manage certain key risks effectively (e.g., mitigating collateral volatility with fiat reserves) while implicitly concentrating others (e.g., custodial and counterparty risks). This risk specialization, evidenced by the ecosystem's near-even split between fiat- and crypto-backed paradigms, creates critical points of failure that often demand centralized governance, challenging the ethos of decentralization.

This landscape of risk specialization is further complicated by a modern market demand: the integration of yield mechanisms. \textbf{This imposes a ``dual mandate'' that systemically breeds new risk.} The integration of yield mechanisms, now a mainstream feature (56.8\% of stablecoins in our study), transforms stablecoins from simple payment tools into complex financial instruments. Fulfilling the mandate for high, competitive returns (with 83.3\% of yield-bearers exceeding the US Treasury benchmark) necessitates high-risk financial engineering, including significant reliance on derivatives and external DeFi protocols. This introduces a fundamental tension between the mission for stability and the strategies required for high returns, creating new vectors for contagion and systemic risk.

When these combined tensions culminate in real-world incidents, the ecosystem’s evolutionary mechanic is laid bare: \textbf{security evolution is forged through trial-by-fire.} Stablecoins undergo a particularly acute evolutionary process driven by security incidents. We find that technical exploits (e.g., code vulnerabilities) and economic attacks that stress-test peg defenses act as stringent ``evolutionary pressures.'' These critical incidents are not merely failures; they are existential tests that necessitate crucial adaptations and redefine the security frontier for subsequent designs.

To translate these analytical insights into a robust evaluative instrument, we introduce the \textbf{Stablecoin LEGO framework}. This quantitative methodology, drawing an analogy from the interlocking toy system, systematically deconstructs past stablecoin failures to their root causes and maps these to identifiable preventive and detective measures within extant implementations. The outcome is a structured, weighted risk score for individual stablecoins. The framework also incorporates the analysis of downstream impacts via token distributions, facilitating a holistic comprehension of stablecoins' pivotal role. Initial application to 11 stablecoins using this framework enables the quantification of disparate risk profiles and reveals how factors, such as the comprehensiveness of security auditing, correlate with diminished assessed risk.

\subsection{Contribution}

Our contributions are summarized as follows:

\begin{enumerate}
    \item A comprehensive systematization of knowledge of the stablecoin landscape: We present the security-focused SoK grounded in a large-scale, multi-source analysis of 157 prior studies, 95 active stablecoins\footnote{Our classification encompasses 95 active stablecoins. Specific sub-analyses may focus on curated subsets for targeted investigation.}, and 44 major security incidents.
    \item Four pivotal insights into stablecoin definition, design, and security: We challenge prevailing assumptions by establishing that stability is an emergent property, design is a trade-off in risk specialization, yield creates a dual mandate with systemic risk implications, and security evolution is driven by critical failures.
    \item The Stablecoin LEGO framework: We propose a novel quantitative methodology for evaluating stablecoin risk by systematically mapping historical failure modes to preventive and detective measures. This enables structured, repeatable risk assessment and supports continuous ecosystem monitoring.
\end{enumerate}

\begin{figure}[!tb]
    \centering
    \includegraphics[width=\linewidth]{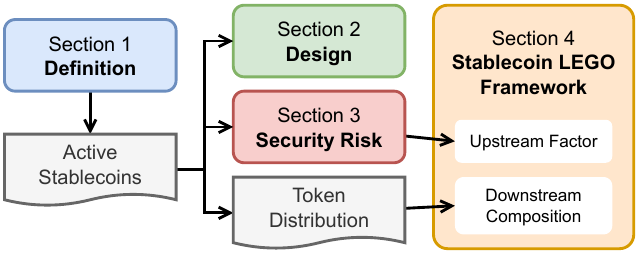}
    \caption{The SoK architecture.}
    \label{fig:sok}
\end{figure}

The SoK architecture is organized in Fig.~\ref{fig:sok}. Specifically, we examine previous studies for stablecoin definitions in Section~\ref{sec:definition}, designs including collateral assets, stabilization mechanisms, and yield mechanisms in Section~\ref{sec:classification}, security risks in Section~\ref{sec:security_risk}, and the Stablecoin LEGO framework in Section~\ref{sec:stablecoin_lego}.
All source code and calculation details are published here~\footnote{\url{https://github.com/stablecoin-sok}}.

\section{Definition}
\label{sec:definition}

Establishing a clear definition of ``stablecoin'' is foundational to systematizing its security landscape. This section delineates the research scope of this paper by dissecting how stablecoins are characterized across academic, governmental, and industry literature. From this comprehensive review, we derive several noteworthy findings regarding the nature and perception of stablecoins.

\subsection{Methodology}
\label{sec:def_methodology}
Our selection criteria targeted a diverse range of research sources to capture a holistic understanding of stablecoins:

\begin{itemize}
    \item \textbf{Academia (Google Scholar \& Top Conferences):} We analyzed the top 100 Google Scholar results for ``stablecoin'' (with over 20 citations) and relevant papers from the last five years published in 34 leading academic conferences across security, privacy, cryptography, networking, database, software engineering, programming language, and system architecture~\cite{conferences}.
    \item \textbf{Governmental \& Intergovernmental Bodies:} We reviewed reports from the past five years issued by G20 member states' financial authorities (e.g., central banks) and key international financial organizations (i.e., IMF, WB, BIS, FSB, FATF). Reports expressing non-official views were excluded for rigor\footnote{We exclude the studies that, although published by certain institutions, have specially claimed irrelevance to official views. For instance, some IMF reports claim that ``the views expressed in Fintech Notes are those of the author(s) and do not necessarily represent the views of the IMF, its Executive Board, or IMF management.''}.
    \item \textbf{Industry (from Web3 Media):} We examined stablecoin-related articles and news from the past five years from the top 5 Web3 media outlets (Cointelegraph, CoinDesk, BeInCrypto, Crypto News, Decrypt), identified via web traffic metrics (details in Appendix~\ref{sec:web3_media}).
\end{itemize}

\subsection{Result and Findings}

This methodology yielded 157 research studies (56 academic, 81 governmental, 20 industry-focused), the definitions from which are summarized in Appendix~\ref{app:definition}. Our analysis of this corpus reveals several key insights into the evolving understanding of stablecoins.

\subsubsection{Finding 1: ``Stablecoin'' is a Contested and Developing Term}

A primary finding, consistently highlighted in governmental and regulatory literature, is that ``stablecoin'' remains an evolving, collective term lacking a universally agreed-upon technical definition~\cite{fsb_2022_2,fsb_2023_3}. Crucially, the term itself is not an affirmation of achieved stability but is often employed as a marketing label by market participants and authorities~\cite{fatf_2020_2,fsb_2022_2,fsb_2023_3}. Consequently, these assets may not always maintain their peg and can exhibit risk profiles comparable to other volatile cryptoassets~\cite{g7_2019}.

\subsubsection{Finding 2: Stablecoin Research Exhibits a Blooming Trend with Shifting Focus}

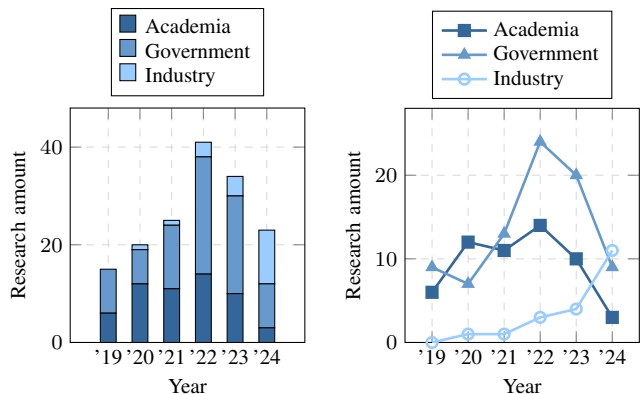
\begin{figure}[!tb]
    \footnotesize
    \begin{subfigure}[b]{0.24\textwidth}
        \begin{tikzpicture}
            \begin{axis}[
                width=1.08\textwidth,
                height=1.08\textwidth,
                ybar stacked, 
                bar width=6pt, 
                symbolic x coords={'19, '20, '21, '22, '23, '24}, 
                xtick=data, 
                ymin=0, 
                ymax=48, 
                xlabel={Year}, 
                ylabel={Research amount}, 
                legend style={
                    at={(0.5,1.05)}, 
                    anchor=south, 
                    legend columns=1, 
                    column sep=0.05cm,
                    row sep=-0.05cm,
                },
                grid=both,
                grid style={dashed,gray!30},
                legend cell align={left},
                enlarge x limits=0.24, 
            ]
            \addplot[fill=2] coordinates {('19,6) ('20,12) ('21,11) ('22,14) ('23,10) ('24,3)};
            \addplot[fill=3] coordinates {('19,9) ('20,7) ('21,13) ('22,24) ('23,20) ('24,9)};
            \addplot[fill=4] coordinates {('19,0) ('20,1) ('21,1) ('22,3) ('23,4) ('24,11)};
            \legend{Academia, Government, Industry}
            \end{axis}
        \end{tikzpicture}
    \end{subfigure}
    % \hfill
    \begin{subfigure}[b]{0.23\textwidth}
        \begin{tikzpicture}
            \begin{axis}[
                width=1.126\textwidth,
                height=1.126\textwidth,
                xlabel={Year},
                ylabel={Research amount},
                symbolic x coords={'19, '20, '21, '22, '23, '24}, 
                xtick=data, 
                ymin=0,
                ymax=28,
                % legend pos=north west,
                legend style={
                    at={(0.5,1.04)}, 
                    anchor=south, 
                    legend columns=1, 
                    column sep=0cm,
                    row sep=-0.05cm,
                },
                grid=both,
                grid style={dashed,gray!30},
                ylabel near ticks,
                xlabel near ticks,
                legend cell align={left},
                enlarge x limits=0.15, 
            ]
            \addplot[line width=1pt, color=2, mark=square*] coordinates {('19,6) ('20,12) ('21,11) ('22,14) ('23,10) ('24,3)};
            \addlegendentry{Academia}
            \addplot[line width=1pt, color=3, mark=triangle*] coordinates {('19,9) ('20,7) ('21,13) ('22,24) ('23,20) ('24,9)};
            \addlegendentry{Government}
            \addplot[line width=1pt, color=4, mark=o] coordinates {('19,0) ('20,1) ('21,1) ('22,3) ('23,4) ('24,11)};
            \addlegendentry{Industry}
            \end{axis}
        \end{tikzpicture}
    \end{subfigure}
    \caption{The year trend of the amount of prior stablecoin research.}
    \label{fig:research_trend}
\end{figure}

As illustrated in Fig.~\ref{fig:research_trend}, the volume of research publications surged from 2019, peaking in 2022. Notably, while academic and governmental research output saw a subsequent decline, industry-focused research and analysis appear to maintain a rising trajectory. This divergence might suggest a recalibration period post-2022 (coinciding with major stablecoin failures), with regulators and academics perhaps adopting a more cautious, observational stance, while the industry continues to innovate and explore new models, potentially driven by persistent market demand or a search for more resilient designs.

\subsubsection{Finding 3: Definitional Diversity Underscores Stablecoins as Adaptive Systems}
\label{sec:diverse_definitions}

Analysis of the collected definitions (Fig.~\ref{fig:definition}) reveals significant heterogeneity across three key descriptive dimensions:

\begin{figure}[!tb]
    \footnotesize
    \begin{subfigure}[b]{0.4\textwidth}
    \begin{tikzpicture}
        \begin{axis}[
            xbar stacked, 
            bar width=6pt, 
            symbolic y coords={Stability requirement, Pegged asset, Underlying platform},
            ytick=data,
            xmin=0, 
            xmax=165, 
            width=\textwidth, 
            height=0.4\textwidth, 
            enlarge y limits=0.46, 
        ]
        \addplot[xbar, fill=1] coordinates {(11,Underlying platform)(0,Pegged asset)(0,Stability requirement)};
        \addplot[xbar, fill=2] coordinates {(9,Underlying platform)(0,Pegged asset)(0,Stability requirement)};
        \addplot[xbar, fill=blue3] coordinates {(7,Underlying platform)(0,Pegged asset)(0,Stability requirement)};
        \addplot[xbar, fill=blue4] coordinates {(1,Underlying platform)(0,Pegged asset)(0,Stability requirement)};
        \addplot[xbar, fill=blue5] coordinates {(130,Underlying platform)(0,Pegged asset)(0,Stability requirement)};
        \addplot[xbar, fill=red1] coordinates {(0,Underlying platform)(113,Pegged asset)(0,Stability requirement)};
        \addplot[xbar, fill=red2] coordinates {(0,Underlying platform)(23,Pegged asset)(0,Stability requirement)};
        \addplot[xbar, fill=red3] coordinates {(0,Underlying platform)(22,Pegged asset)(0,Stability requirement)};
        \addplot[xbar, fill=green1] coordinates {(0,Underlying platform)(0,Pegged asset)(85,Stability requirement)};
        \addplot[xbar, fill=green2] coordinates {(0,Underlying platform)(0,Pegged asset)(30,Stability requirement)};
        \addplot[xbar, fill=green3] coordinates {(0,Underlying platform)(0,Pegged asset)(25,Stability requirement)};
        \addplot[xbar, fill=green4] coordinates {(0,Underlying platform)(0,Pegged asset)(7,Stability requirement)};
        \addplot[xbar, fill=green5] coordinates {(0,Underlying platform)(0,Pegged asset)(11,Stability requirement)};
        \end{axis}
    \end{tikzpicture}
    \end{subfigure}
    % \vskip 0.2cm
    \begin{subfigure}[b]{0.4\textwidth}
        \centering
        \begin{tikzpicture}
            \node[draw, fill=1, minimum width=0.2cm, minimum height=0.2cm] (up1) at (0, 0) {}; 
            \node[right=0.1cm of up1] {DLT};
            \node[draw, fill=2, minimum width=0.2cm, minimum height=0.2cm] (up2) at (1.3, 0) {};
            \node[right=0.1cm of up2] {Blockchain};
            \node[draw, fill=blue3, minimum width=0.2cm, minimum height=0.2cm] (up3) at (3.3, 0) {};
            \node[right=0.1cm of up3] {Public blockchain};
            \node[draw, fill=blue4, minimum width=0.2cm, minimum height=0.2cm] (up4) at (6.2, 0) {};
            \node[right=0.1cm of up4] {Crypto. tech.};
            \node[draw, fill=blue5, minimum width=0.2cm, minimum height=0.2cm] (up5) at (6.5, -0.35) {};
            \node[right=0.1cm of up5] {Unspecified};

            \node[draw, fill=red1, minimum width=0.2cm, minimum height=0.2cm] (pa1) at (0, -0.35) {}; 
            \node[right=0.1cm of pa1] {The specified};
            \node[draw, fill=red2, minimum width=0.2cm, minimum height=0.2cm] (pa2) at (2.3, -0.35) {};
            \node[right=0.1cm of pa2] {Fiat currency};
            \node[draw, fill=red3, minimum width=0.2cm, minimum height=0.2cm] (pa3) at (4.4, -0.35) {};
            \node[right=0.1cm of pa3] {Unspecified};

            \node[draw, fill=green1, minimum width=0.2cm, minimum height=0.2cm] (sr1) at (0, -0.7) {}; 
            \node[right=0.1cm of sr1] {Stable};
            \node[draw, fill=green2, minimum width=0.2cm, minimum height=0.2cm] (sr2) at (1.4, -0.7) {};
            \node[right=0.1cm of sr2] {Peg/tie/link/track};
            \node[draw, fill=green3, minimum width=0.2cm, minimum height=0.2cm] (sr3) at (4.1, -0.7) {};
            \node[right=0.1cm of sr3] {Minimize/lower/low/less/mitigate};
            \node[draw, fill=green4, minimum width=0.2cm, minimum height=0.2cm] (sr4) at (0, -1.05) {};
            \node[right=0.1cm of sr4] {Fixed/constant or close/near-constant/near-fixed};
            \node[draw, fill=green5, minimum width=0.2cm, minimum height=0.2cm] (sr5) at (6.5, -1.05) {};
            \node[right=0.1cm of sr5] {Unspecified};
        \end{tikzpicture}
    \end{subfigure}
    
    \caption{Stablecoin definitions in terms of underlying platform, pegged asset, and stability requirement.}
    \label{fig:definition}
\end{figure}
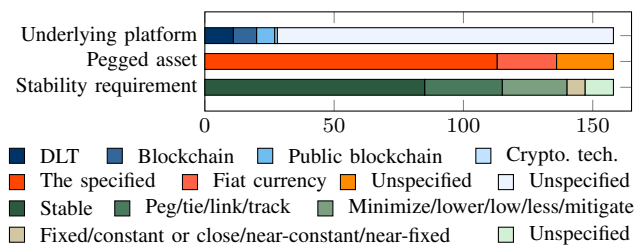

\begin{itemize}
    \item \textbf{Underlying Platform:} A vast majority of definitions (82.28\%) do not specify a particular platform type. Those that do describe a spectrum from general Distributed Ledger Technology (DLT) to more specific ``blockchain'' or ``public blockchain'' technologies.
    \item \textbf{Pegged Asset:} Most definitions (71.52\%) require a specified reference asset (which can include fiat currency, real-world assets (RWAs), or other cryptoassets). A smaller subset (14.56\%) restricts this peg exclusively to fiat currencies, while the remainder (13.92\%) lack specificity.
    \item \textbf{Stability Requirement:} There is little consensus here. Approximately half (53.80\%) merely use the term ``stable'', implying a desired state rather than a strict technical criterion. Others employ verbs like ``peg/tie/link/track'' (18.99\%) or aim to ``minimize/lower/mitigate'' volatility (15.82\%). Only a small fraction (4.43\%) explicitly demand a fixed or near-fixed value, with the rest (6.96\%) not detailing the stability criterion.
\end{itemize}

The inherent vagueness in these common definitional components, particularly regarding the ``stability requirement'', suggests an implicit acceptance of potential price fluctuations. As noted by the Deutsche Bundesbank~\cite{de_2022}, the price of a stablecoin is not perfectly correlated with its reference asset due to supply and demand dynamics on trading platforms. This underscores a crucial nature: stablecoins are better understood as adaptive socio-technical systems rather than static monetary instruments. Their stability is consequently a dynamic and often fragile equilibrium, not an inherent, guaranteed property.

\subsection{Research Scope}
Given the definitional landscape, we establish our research scope for this SoK as follows:

\noindent\textbf{Broad definition.}
We acknowledge the widely accepted definition from the Financial Stability Board (FSB)~\cite{fsb_2019}, entrusted by the G20: ``\emph{A crypto-asset that aims to maintain a stable value relative to a specified asset, or a pool or basket of assets.}''

\noindent\textbf{Strict definition.}
While the FSB definition is encompassing, significant regulatory and systemic risk concerns prioritize stablecoins pegged to fiat currencies. These are perceived to have a greater potential to become widely accepted means of payment, thereby posing more immediate and substantial monetary and financial stability risks~\cite{hk_2024_2,fsb_2024}. Therefore, for the purpose of this SoK, we adopt a strict definition: ``\emph{A crypto-asset that aims to maintain a stable value relative to a specified fiat currency, or a pool or basket of fiat currencies.}'' This focused scope allows for a deeper and more coherent analysis of the security risks pertinent to the most systemically relevant class of stablecoins.

\subsection{Similar Concepts}
It is also noteworthy that analogous concepts exist within the aforementioned stablecoin definitions which are prone to confusion. However, in this paper, we intentionally exclude these concepts, as they are either deliberately or incidentally developed for distinct purposes and fall outside the scope of the established community consensus on stablecoins.

\noindent\textbf{Central Bank Digital Currency (CBDC).}
CBDC is usually the digital form of central bank currency instead of third parties, and may or may not adopt technologies like distributed ledger or blockchain. CBDC would create a modern alternative to stablecoins, as suggested by Deutsche Bundesbank~\cite{de_2022}.

\noindent\textbf{Tokenized fund.} A digital representation of an asset or ownership right as a token on a blockchain~\cite{tokenized_fund_Deloitte,tokenized_fund_kpmg,tokenized_fund_ey,tokenized_fund_pwc}, exemplified by Franklin Templeton FOBXX~\cite{fobxx} and BlackRock BUIDL~\cite{buidl}. This concept is excluded from this paper as it aligns more closely with financial constructs regulated by securities laws and primarily caters to the asset management and investment sectors, and should be viewed as an alternative to secured stablecoins or a supplement to CBDCs suggested by the Bank of Russia~\cite{ru_2023}.
    
\noindent\textbf{Wrapped token.} A digital asset that reflects the value of another cryptocurrency from a different blockchain, such as Wrapped BTC (WBTC)~\cite{wbtc} on Ethereum, aiming at addressing the challenge of interoperability across blockchains~\cite{wrapped_coinbase,wrapped_binance,info13010006}. This concept is excluded from this paper as it primarily functions as an interoperability solution rather than maintaining value stability.

\noindent\textbf{Bridged token.} A digital asset that is bridged from one blockchain to the other via a cross-chain bridge. Typical examples include USDC (Ethereum) - USDC.e (Optimism)~\cite{usdce}. It differs from a wrapped token in that it may have already natively existed on the target blockchain before bridging, while still excluded for the same reason. 

\noindent\textbf{Liquidity provider (LP) token.} A token issued to liquidity providers on AMM protocols, tracking individual shares to the overall liquidity pool~\cite{lp_cmc,lp_binance}. We exclude it because it primarily exists within the AMM system as an ownership certificate, which can take other forms, such as NFTs.
    
\noindent\textbf{Liquidity staking token (LST).} Also known as liquidity staking derivative (LSD), tokenized representations of staked tokens~\cite{gogol2024sokliquidstakingtokens,scharnowski2024economics}. Typical examples include Lido stETH~\cite{steth} on Ethereum. We exclude it from this paper because they are considered add-on derivatives of liquidity staking.

\insight{Distinct from multi-utility DeFi tokens, a stablecoin's sole objective is peg stability. As adaptive socio-technical systems, this vital function fundamentally relies on two vital, market-validated conditions: 1) sustained market confidence, paramount due to absent universal backing rules and earned through transparent collateral and robust stabilization, and 2) effective convertibility (liquidity) ensuring consistent exchange for its reference value. }

\section{Design}
\label{sec:classification}

Building upon the strict definition of stablecoins (Section~\ref{sec:definition}), this section deconstructs their design landscape. While a common initial categorization distinguishes between collateralized stablecoins (backed by assets) and algorithmic stablecoins (relying on dynamic mechanisms), this distinction is not absolute. Many stablecoins employ hybrid approaches, combining collateralization with algorithmic adjustments to pursue stability. We therefore adopt a multi-faceted approach to classify and analyze stablecoin designs.

\subsection{Methodology}
\label{sec:stablecoin_selection}

To systematically understand stablecoin design, we identify three primary attributes as classification criteria: 1) Collateral Asset types, 2) Stabilization Mechanisms, and 3) native Yield Mechanisms. Our analysis covers 95 existing stablecoins, selected by market capitalization (over \$10M from sources of DefiLlama~\cite{defillama_stable}, CoinMarketCap~\cite{coinmarketcap_stable}, and CoinGecko~\cite{coingecko_stable}). We verified features via official documentation and excluded failed or inactive projects to focus on currently operational designs (see Appendix~\ref{app:existing_stablecoins} for the list).

\begin{figure}[!tb]
    \centering
    \includegraphics[width=\linewidth]{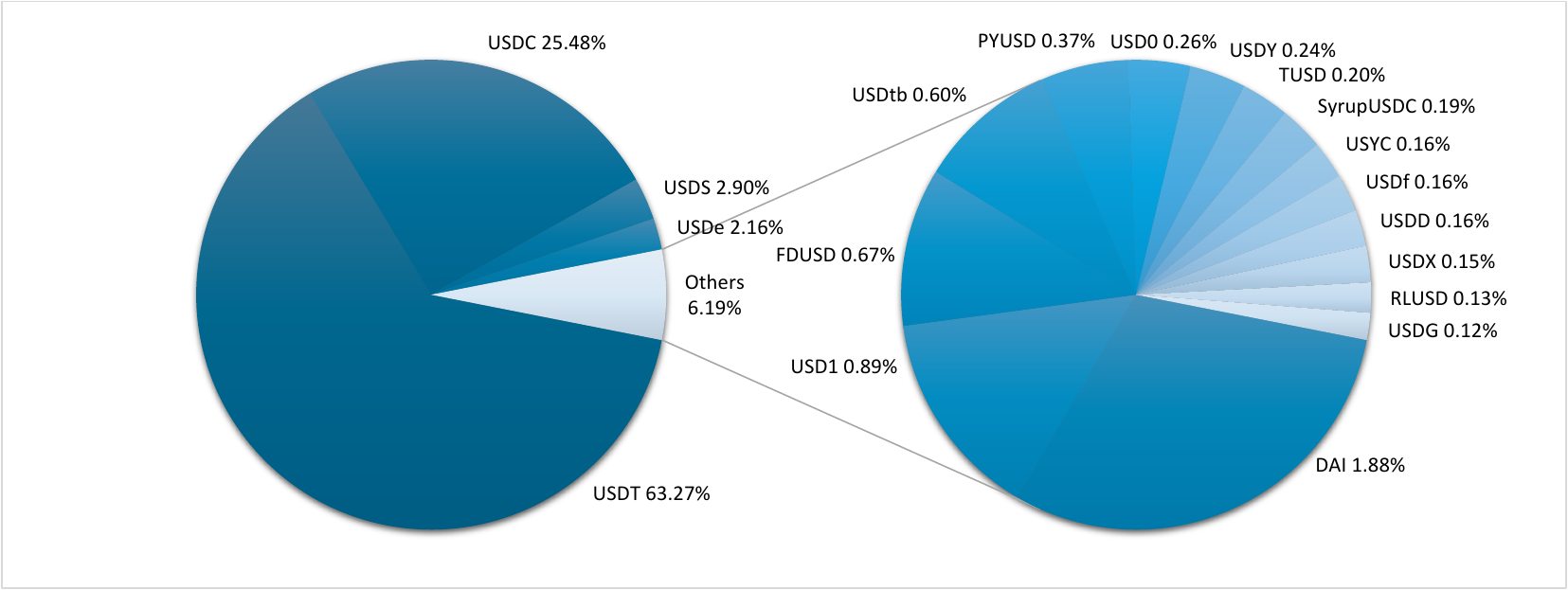}
    \caption{The distribution of top 20 stablecoins regarding market capitalization (in million USD).}
    \label{fig:top20}
\end{figure}

\noindent\textbf{Observation 1: Market Concentration.} The stablecoin market is highly concentrated: the top 5 (USDT, USDC, USDS, USDe, DAI) constitute over 93\% of total market capitalization, and the top 20 represent 98\% (Fig.~\ref{fig:top20}). This underscores the dominance of a few major stablecoins, consistent with the Pareto Principle.

\noindent\textbf{Observation 2: Motivations for Stablecoin Emergence.} Despite market concentration, new stablecoins continually emerge, driven by diverse motivations beyond simple price stability:
\begin{itemize}
    \item Regional demand: catering to local economies with fiat-pegged stablecoins (e.g., EURS for Euro).
    \item Ecosystem demand: providing native stablecoins for burgeoning blockchain ecosystems (e.g., Blast USDB).
    \item Decentralization focus: offering alternatives (e.g., MakerDAO DAI) to centralized issuers like Tether, aiming to mitigate counterparty risks.
    \item Stability innovation: introducing novel mechanisms, e.g., hedging strategies, to enhance price stability (e.g., Ethena USDe).
    \item Financial innovation: incorporating new economic models, governance structures, or yield-bearing features (e.g., Sky USDS).
\end{itemize}

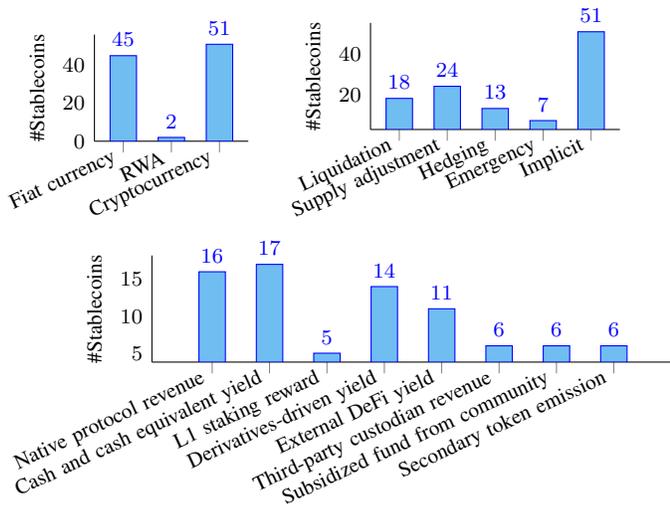
\begin{figure}[!tb]
  \centering
  \begin{subfigure}[b]{0.2\textwidth}
  \begin{tikzpicture}
    \begin{axis}[
      ybar,
      ymin=0,
      bar width=10pt,
      width=1\linewidth,
      height=3cm,
      ylabel style={font=\footnotesize},
      ylabel={\#Stablecoins},
      symbolic x coords={
        Fiat currency,
        RWA,
        Cryptocurrency
      },
      xtick=data,
      x tick label style={rotate=25, anchor=east, font=\footnotesize},
      ytick style={draw=none},
      axis x line*=bottom,
      axis y line*=left,
      enlarge x limits=0.3, 
      tick label style={font=\footnotesize},
      every node near coord/.append style={font=\footnotesize, rotate=0, anchor=south},
      nodes near coords
    ]
    \addplot+[ybar, fill=blue3] coordinates {
        (Fiat currency,45)
        (RWA,2)
        (Cryptocurrency,51)
    };
    \end{axis}
  \end{tikzpicture}
  \end{subfigure}
  \begin{subfigure}[b]{0.27\textwidth}
      \begin{tikzpicture}
      \begin{axis}[
        ybar,
        bar width=10pt,
        width=1\linewidth,
        height=3cm,
        ylabel style={font=\footnotesize},
        ylabel={\#Stablecoins},
        symbolic x coords={
          Liquidation,
          Supply adjustment,
          Hedging,
          Emergency,
          Implicit
        },
        xtick=data,
        x tick label style={rotate=25, anchor=east, font=\footnotesize},
        ytick style={draw=none},
        axis x line*=bottom,
        axis y line*=left,
        enlarge x limits=0.15,
        tick label style={font=\footnotesize},
        every node near coord/.append style={font=\footnotesize, rotate=0, anchor=south},
        nodes near coords
      ]
      \addplot+[ybar, fill=blue3] coordinates {
          (Liquidation,18)
          (Supply adjustment,24)
          (Hedging,13)
          (Emergency,7)
          (Implicit,51)
      };
      \end{axis}
    \end{tikzpicture}
  \end{subfigure}
  \begin{subfigure}[b]{0.47\textwidth}
    \begin{tikzpicture}
      \begin{axis}[
        ybar,
        bar width=10pt,
        width=1\linewidth,
        height=3cm,
        ylabel style={font=\footnotesize},
        ylabel={\#Stablecoins},
        symbolic x coords={
          Native protocol revenue,
          Cash and cash equivalent yield,
          L1 staking reward,
          Derivatives-driven yield,
          External DeFi yield,
          Third-party custodian revenue,
          Subsidized fund from community,
          Secondary token emission
        },
        xtick=data,
        x tick label style={rotate=25, anchor=east, font=\footnotesize},
        ytick style={draw=none},
        axis x line*=bottom,
        axis y line*=left,
        enlarge x limits=0.15,
        tick label style={font=\footnotesize},
        every node near coord/.append style={font=\footnotesize, rotate=0, anchor=south},
        nodes near coords
      ]
      \addplot+[ybar, fill=blue3] coordinates {
          (Native protocol revenue,16)
          (Cash and cash equivalent yield,17)
          (L1 staking reward,5)
          (Derivatives-driven yield,14)
          (External DeFi yield,11)
          (Third-party custodian revenue,6)
          (Subsidized fund from community,6)
          (Secondary token emission,6)  
      };
      \end{axis}
    \end{tikzpicture}
  \end{subfigure}
  \caption{Stablecoin distributions in terms of collateral asset, stabilization mechanism, and yield sources (across yield-bearing stablecoins).}
  \label{fig:classification}
\end{figure}

\subsection{Collateral Asset}
\label{sec:collateral}

Collateral assets are fundamental to many stablecoin designs, underpinning their purported value. Before analysis, we clarify crucial distinctions: the pegged asset is the target value (e.g., USD); the collateral asset backs the stablecoin; and the purchase fund is the medium for acquiring the stablecoin, not necessarily linked to its peg or collateral.
We categorize collateral into: 1) \textit{Fiat currency} (or equivalents like government treasuries), 2) \textit{Real-World Assets (RWA)} (e.g., commodities, real estate, stocks), and 3) \textit{Cryptocurrency} (e.g., USDT, BTC, ETH).

\noindent\textbf{Collateralization Landscape.} Our analysis of 95 operational stablecoins reveals that all are, at least nominally, fully collateralized (i.e., collateral value larger or equal to 100\% of outstanding supply). This suggests a strong market tendency towards, or higher survival rate for, designs with explicit full backing, where complex stabilization algorithms often act as secondary or reinforcing measures. Among these, 45 are primarily fiat-backed, 2 by RWA, and 51 by cryptocurrencies (total exceeds 95 due to multi-collateral designs), indicating a near-even split in preference between fiat and crypto-collateral paradigms (Fig.~\ref{fig:classification}).

\noindent\textbf{Comparative Analysis of Collateral Types.} 
We evaluated USD (fiat), Gold (RWA), and Bitcoin (cryptocurrency) against four key attributes (Table~\ref{tab:collateral_benefit_risk}), including volatility, redemption efficiency, inflation resistance, and compliance. We recognize that liquidity is paramount, with inflation and compliance risks impacting this core tenet.
This comparison reveals inherent trade-offs: each asset type excels in some dimensions while underperforming in others, underscoring that collateral choice fundamentally dictates a stablecoin's risk-return profile and operational characteristics.

\begin{table*}[!tb]
    \centering
    \begin{tabular}{c|cccc}
        \toprule
         & Volatility (PSD) & Redemption efficiency (REI) & Inflation resistance ($r$)& Compliance (J-Score) \\
        \midrule
        USD (fiat currency) & \textbf{5.93} & 0.9990 & -4.25 & \textbf{21} \\  % 0-4.25
        Gold (RWA) & 313.77 & 2.0000 & \textbf{5.89} & \textbf{21} \\  % 5.89-0
        Bitcoin (cryptocurrency) & 23413.08 & \textbf{0.0000} & \textbf{5.78} & 13 \\  %5.78-0
        \bottomrule
    \end{tabular}
    \caption{Comparison of three collateral assets, where bold numbers are better ones. }
    \label{tab:collateral_benefit_risk}
\end{table*}

\subsubsection{Volatility} 

Collateral asset volatility directly impacts a stablecoin's ability to maintain its peg. Highly volatile collateral necessitates more aggressive stabilization mechanisms and can undermine user confidence. While fiat currencies, particularly the USD, are generally regarded as the most price-stable collateral options, RWAs like gold exhibit moderate fluctuation based on market dynamics. Cryptocurrencies represent the most volatile class, with potential for drastic price swings.
To quantify this, we use the Price Standard Deviation (PSD):
\begin{align}
PSD = \sqrt{\frac{1}{T} \sum_{t=1}^{T} (P(t) - \mu)^2},
\end{align}
where $T$ is the observation period, $P(t)$ is the asset price at time $t \in T$, and $\mu$ is the mean price over $T$. We calculated PSD using daily closing prices from Yahoo Finance for the five-year period from March 25, 2020, to March 24, 2025. Our analysis confirms that USD exhibits the lowest volatility, whereas Bitcoin demonstrates the highest among the three representative assets.

\subsubsection{Redemption Efficiency}

Redemption efficiency, which is the ease and speed with which collateral can be converted to meet redemption demands without adverse price impact, is crucial for stablecoin trustworthiness. Fiat currencies offer high global liquidity and accessibility. RWAs can face logistical hurdles and slower conversion times. Cryptocurrencies present variable liquidity dependent on the specific asset and prevailing market conditions, potentially stressing stability during demanding redemption periods.
We evaluate this using a Redemption Efficiency Index (REI), grounded in market microstructure theory, which amalgamates normalized transaction costs and redemption delays:
\begin{equation}
    REI = f' + d',
\end{equation}
where $f',d'$ are min-max normalized values (scaled to [0,1]) representing typical transaction fees (in USD equivalents) and redemption delays (in days) associated with converting the collateral asset. A lower REI signifies higher efficiency. Our analysis indicates Bitcoin offers the highest redemption efficiency due to its near-instant, on-chain settlement capabilities, while physical gold ranks lowest due to logistical requirements. USD efficiency is high but typically subject to banking system operational hours and settlement lags.

\subsubsection{Inflation Resistance} 

Inflation erodes the purchasing power of assets used as collateral. Assets that can hedge against inflation are therefore valuable for preserving a stablecoin's real value. While certain cryptocurrencies, particularly those with capped supplies, are posited as inflation hedges, fiat currencies directly lose purchasing power during inflationary periods. RWAs like gold have a mixed historical record as consistent inflation hedges.

We assess inflation resistance using the real return ($r$), derived from the Fisher Equation: $(1+i)=(1+r)(1+\pi)$, where $i$ is the nominal return of the collateral asset and $\pi$ is the relevant annual inflation rate. For low inflation, this approximates to: 
\begin{equation}
    r \approx i - \pi .
\end{equation} 
We determine $i$ by taking the median yield rate offered by stablecoins in our dataset that are collateralized by the respective asset type (e.g., median yield of USD-backed stablecoins for USD's nominal return). This proxy reflects the returns generated and passed on by stablecoin issuers utilizing that collateral. Our results show that gold and Bitcoin-backed stablecoin models offer superior inflation resistance, while USD-backed models demonstrate negative real returns, reflecting an erosion of purchasing power.

\subsubsection{Compliance} 

The regulatory treatment of collateral assets across jurisdictions introduces significant, often unpredictable, risk. While fiat currencies and traditional RWAs like gold are generally accepted within established regulatory frameworks in most G20 nations, cryptocurrencies navigate a more complex and rapidly evolving legal landscape, impacting their suitability and reliability as collateral.

To quantify this, we propose a Legal and Jurisdictional Compliance Score (J-Score), a primarily qualitative aggregation:
\begin{equation}
J = \sum_{k=1}^{N} w_k \cdot C_k,
\end{equation}
where $N$ is the number of G20 jurisdictions considered (N=21), $w_k$ is the weight for jurisdiction $k$ (here, $w_k=1$ for all, signifying equal weight), and $C_k \in \{0,1\}$ indicates whether the collateral asset type is generally considered compliant (1) or faces significant restrictions/lack of clarity (0) for use in financial instruments or as a reserve asset within that jurisdiction. A higher J-Score indicates broader regulatory acceptance of the collateral type. Our analysis suggests that Bitcoin, as a collateral type, faces compliance ambiguities or restrictions in approximately 40\% of G20 jurisdictions, a higher percentage than for USD or Gold.

\noindent\textbf{Discussion: Stablecoin Compliance.} Beyond collateral, stablecoins also face a rapidly evolving regulatory landscape (e.g., EU's MiCA, frameworks in Singapore, US's STABLE and GENIUS, Hong Kong Stablecoins Ordinance). The security implications of these diverse and emerging regulatory demands warrant continuous investigation.

\subsection{Stabilization Mechanism}

Many stablecoins employ explicit mechanisms to actively defend their peg. Our analysis (Fig.~\ref{fig:classification}) shows that while over half (53.68\%) rely on an ``implicit'' mechanism (primarily trust in the issuer and their reserves), others use active strategies. Specifically, liquidation (18.95\%) and supply adjustment (25.26\%) are prevalent, with hedging (13.68\%) and emergency features (7.37\%) also utilized (total exceeds 100\% as some implement multiple mechanisms).

\subsubsection{Liquidation}
Liquidation mechanisms are foundational to many collateralized stablecoins, enforcing solvency by auctioning off undercollateralized positions. When volatile collateral backing a debt position falls below a predetermined threshold (the liquidation ratio), the system permits liquidators to repay a portion of the debt in exchange for seizing the underlying collateral at a discount~\cite{10.1145/3487552.3487811,10.1007/978-3-662-64331-0_24,10.1007/978-3-031-47754-6_20}. This process inherently relies on over-collateralization, where the initial collateral value significantly exceeding the minted stablecoin value, to buffer against price declines. MakerDAO DAI is a prominent example.

\begin{algorithm}[!tb]
    \caption{Liquidation}
    \KwIn{current\_value, debt, liquidation\_threshold, discount, liquidation\_rate}
    \KwOut{seized or "Safe"}
    \If{$current\_value / debt < liquidation\_threshold$}{
        $seized \gets current\_value \times (1 - discount)$\;
        repay($debt \times liquidation\_rate$)\;
        \Return{$seized$}
    }
    \Return{``Safe''}
\end{algorithm}

Considered a relatively robust approach, liquidation is widely adopted. MakerDAO DAI, one of the largest stablecoins employing this, has navigated market volatility without catastrophic failures of its core liquidation engine. A key insight from Klages-Mundt et al.~\cite{10.1145/3419614.3423261} is that such mechanisms shift risk from a systemic ``equity risk'' (borne by all token holders) to an ``agent risk'' (borne by individual, over-collateralized vault owners). This design shares structural similarities with borrowing/lending protocols, facilitating their natural integration or evolution into stablecoin issuers (e.g., Aave GHO).

\subsubsection{Supply Adjustment}
This class of mechanisms aims to stabilize price by algorithmically modulating the stablecoin's circulating supply, based on the economic principle that decreasing supply raises prices and vice-versa. While methods vary from direct minting/burning of tokens to adjusting borrowing interest rates to influence demand, they typically rely on arbitrage incentives for market participants. USDD is a notable example.

Theoretically, such mechanisms find grounding in concepts like the Quantity Theory of Money: $M \cdot V = P \cdot Q$, where adjusting money supply $M$ is intended to influence the price level $P$~\cite{ito2020stablecoin}. By algorithmically contracting supply when the price is below peg (to induce scarcity) or expanding it when above peg (to reduce premium), these systems attempt to dynamically manage inflationary or deflationary pressures on the stablecoin's value.

\begin{algorithm}[!tb]
    \caption{Supply Adjustment}
    \KwIn{current\_supply, current\_price, target\_price, adjustment\_coefficient}
    \KwOut{None}
    $supply\_change \gets current\_supply \times adjustment\_coefficient \times (current\_price - target\_price)$\;
    \If{$current\_price > target\_price$}{
        mint(supply\_change)\;
    }
    \Else{
        burn(abs(supply\_change))\;
    }
\end{algorithm}

In practice, however, stablecoins primarily reliant on endogenous supply adjustments have a notable history of failure (e.g., Terra UST, Neutrino USDN, Beanstalk BEAN, Haven xUSD). A critical lesson from these incidents is their acute susceptibility to reflexive market dynamics and oracle unreliability. The adjustment processes can exhibit significant lags, failing to respond adequately to rapid sentiment shifts or well-capitalized attacks, creating death spirals. The specific security risks inherent in these designs are further explored in Section~\ref{sec:security_risk}.

\subsubsection{Hedging}
Hedging strategies aim to neutralize the price risk of volatile collateral by establishing offsetting positions in derivative markets~\cite{CLEWLOW19971353,0b9b8115-a8b8-3422-8e1c-a62077de6621}. Delta-hedging, for example, seeks a ``delta-neutral'' position where the stablecoin issuer's net exposure to the collateral's price movement is theoretically zero. If minting a stablecoin against 1 ETH creates a +1 ETH price exposure (positive delta), a corresponding short position in an ETH perpetual contract of equivalent size would be taken to neutralize this. Ethena USDe and Elixir deUSD exemplify this approach.

A critical aspect of these designs is their operational dependency on centralized exchanges (CEXs) for providing the necessary derivative instruments and liquidity for hedging. This introduces significant counterparty risk: the failure or compromise of a CEX partner could jeopardize the hedge and, consequently, the stablecoin's backing. Furthermore, for models like Ethena, where yield is partly generated from derivative positions (e.g., funding rates, basis spreads), the sustainability and security of these yields are also contingent on the continuous, reliable functioning of these CEXs, creating layers of external dependencies.

\subsubsection{Emergency Mechanism}
Acknowledging the limitations of purely algorithmic responses in extreme scenarios, some stablecoins incorporate emergency mechanisms. These can include features to temporarily suspend core functions like transfers or redemptions during severe market dislocations or security incidents (e.g., Curve crvUSD, Paxos USDP). Another approach involves maintaining segregated bailout reserves, deployable via a trusted committee or governance vote to recapitalize the system during a crisis (e.g., Gyroscope GYD, dForce USX).

While intended as safety nets, such mechanisms inherently introduce centralization and governance risks. The power to invoke emergency actions often resides with a core team or a small group of token holders, raising concerns about potential abuse, or failure to act appropriately under pressure. These mechanisms shift trust from fully autonomous code to human judgment and intervention, introducing governance challenges that extend beyond typical smart contract rule-based security considerations.

\subsubsection{Implicit Stabilization Mechanism}
This dominant category (51/95) includes stablecoins lacking explicit, algorithmically-defined stabilization protocols, relying instead on users' trust in the issuer's ability and commitment to maintain the peg, typically through robust reserve management and redemption processes (e.g., USDT, USDC). The ``stability'' here is an emergent property of this trust and the perceived strength of off-chain backing. However, this category exhibits extremes: from highly resilient, large-cap stablecoins to smaller, more vulnerable ones where ``implicit'' may equate to an absence of robust defenses, heightening risks like rug pulls if issuer trust is misplaced.

\subsubsection{Discussion: Towards a Control-Theoretic View of Stabilization}

No single stabilization mechanism is flawless, thus effective peg maintenance often requires judicious strategy selection and scaling. 
Control theory finds extensive applications in financial markets, including monetary policy control, portfolio optimization, trading and market making, and price stabilization for exchange rates and commodities.
Therefore, we propose that stablecoin stabilization can be fruitfully modeled as a control system. Price deviations from the peg act as error signals, triggering corrective actions from stabilization mechanisms (control inputs $u$) to counteract disturbances $d$ (e.g., market volatility) and guide the system state $x$ (stablecoin price $P(t)$) back to its target $P_{peg}$. A general state-space representation could be: 
\begin{equation}
    dP(t) = ( A P(t) + B U(t) ) dt + \sigma(P(t)) dW(t),
\end{equation}
where $U(t)$ is a vector of control inputs from mechanisms like supply adjustment ($u_s$), liquidation ($u_l$), etc. While a full quantitative development is future work, this control-theoretic perspective offers a powerful abstraction for analyzing the dynamic interactions and potential optimality of combined stabilization strategies.

\insight{The pursuit of stability forces choices that result in \textit{risk specialization rather than comprehensive risk elimination}. This means designs typically manage certain key risks effectively (e.g., reserve volatility in a fiat-backed model) while implicitly accepting or concentrating other risks (like custodial and counter-party risks in the same model), forming a trade-off evidenced by the ecosystem's near-even split between fiat-backed (47.37\%) and crypto-backed (53.68\%) paradigms. This risk concentration creates critical points of failure that, in turn, demand centralized governance for decisive action and accountability, directly challenging the ethos of decentralization. }

\subsection{Yield Mechanism}
\label{sec:yield}

Native yield offerings are a significant driver for stablecoin adoption and innovation, positioning them as financial products beyond simple payment tools. Our analysis considers yields directly provided by issuers, excluding third-party protocol yields, across the 95 selected stablecoins.

\subsubsection{Yield Rate Landscape}

\begin{figure}[!tb]
    \centering
    \footnotesize
    \begin{tikzpicture}
    \begin{axis}[
        width=0.5\textwidth, height=5.5cm,
        xlabel={Market capitalization (million USD)}, ylabel={Yield rate (\%)},
        xmode=log,
        log basis x=10,
        xmin=5, xmax=190000, 
        xtick={10, 100, 1000, 10000, 160000},
        xticklabels={10, 100, 1000, 10000, 160000},
        extra x ticks={50000},
        extra x tick labels={$\cdots$}, 
        extra x tick style={grid=none, tick style={color=white}},
        ymode=log, 
        log basis y=2,
        ymin=1, ymax=90, 
        ytick={2, 4, 8, 16, 32, 64},
        yticklabels={2, 4, 8, 16, 32, 64},
    ]
    \addplot[only marks, mark=*, color=black, mark size = 1.5pt] coordinates {
    (152797,0)
    (61523,0)
    (7007,6.5)
    (5216,4)
    (4539,0)
    (2152,0)
    (1628,0)
    (1443,0)
    (904,0)
    (635,10)
    (580,4.35)
    (494,0)
    (456,10.1)
    (390,0)
    (384,9.4)
    (376,20)
    (373,8.23)
    (310,0)
    (278,0)
    (259,8)
    (250,4.23)
    (244,13.5)
    (237,4.32)
    (235,0)
    (227,3.9)
    (201,5)
    (184,10.72)
    (184,5.79)
    (168,1.1)
    (143,8.63)
    (140,0)
    (137,3.7)
    (137,5.9)
    (126,0)
    (122,15.7)
    (120,21.77)
    (88,0)
    (79,8.75)
    (75,8.5)
    (72,0)
    (68,7.15)
    (65,27.21)
    (62,7.99)
    (60,4.25)
    (58,0)
    (57,0)
    (55,17.68)
    (51,16.93)
    (50,6.86)
    (49,4.4)
    (49,0)
    (48,16.21)
    (48,0)
    (47,4.5)
    (47,0)
    (46,0)
    (46,12.1)
    (42,8.2)
    (41,0)
    (40,0)
    (35,0)
    (34,20)
    (34,0)
    (29,0)
    (28,7.94)
    (27,0)
    (26,0)
    (26,5.53)
    (25,0)
    (24,0)
    (24,11.56)
    (24,0)
    (23,8.46)
    (23,20.57)
    (22,7.38)
    (22,4.31)
    (21,4.3)
    (19,0)
    (18,11)
    (16,9.43)
    (15,8)
    (15,3.99)
    (14,0)
    (14,0)
    (13,4.175)
    (13,0)
    (12,0)
    (11,0)
    (11,14)
    (11,0)
    (11,0)
    (10,0)
    (10,0)
    (10,26.9)
    (10,0)
    };
    \end{axis}
    \end{tikzpicture}
    \caption{The yield distribution of stablecoins paired by market capitalization (log form due to imbalanced data).}
    \label{fig:yield_distribution}
\end{figure}
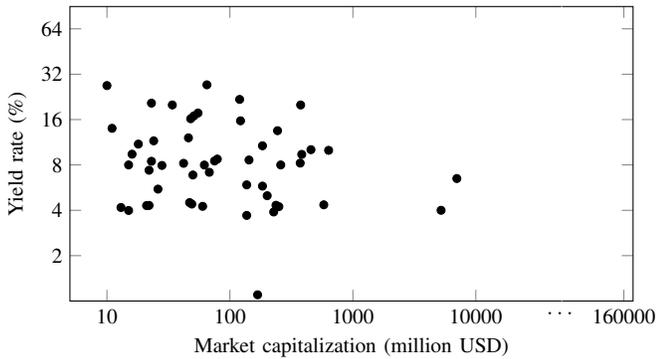

We present the relationship between reported yields and market capitalizations using a log-log distribution (Fig.~\ref{fig:yield_distribution}) for effective visualization. This analysis reveals a distinct inverse correlation: stablecoins with smaller market capitalizations tend to offer higher native yield rates. We posit this reflects a common strategy among emerging or smaller protocols to aggressively attract users and compete for market share, often by offering premium returns to incentivize early adoption.
Quantitatively, our survey of 95 operational stablecoins indicates that 54 (56.84\%) provide native yield-bearing features. Notably, among these yield-bearing stablecoins, 45 (constituting 83.33\% of this subset) offer annual percentage yields (APYs) exceeding 4.25\%. This threshold renders their offerings competitive with, or superior to, traditional benchmarks of the US 10-year Treasury yield (at the time of writing).

\subsubsection{Yield Source Taxonomy and Findings}
We identified 8 primary yield generation patterns: 1) Native protocol revenue, 2) Cash and cash equivalent yield (e.g., T-bills), 3) L1 staking rewards, 4) Derivatives-driven yield (e.g., basis trading, funding rates), 5) External DeFi protocol yield, 6) Third-party custodian revenue, 7) Community-subsidized funds, and 8) Secondary token emissions.
Our analysis (Fig.~\ref{fig:classification}) reveals a complex interplay between traditional financial models and crypto-native risk appetites:

\noindent\textbf{1. Dichotomy in Yield Generation: TradFi Roots with Crypto-Native Risk Layers.} The most prevalent yield sources are anchored in traditional finance: ``Cash and cash equivalent yield'' (utilized by 31.48\% of yield-bearing stablecoins in our study) and ``Native protocol revenue'' from fees (29.63\%). This reliance on established models suggests a market inclination towards perceived sustainability. Yet, in striking contrast, the third most common source is ``Derivatives-driven yield'' (25.93\%), signaling a significant embrace of crypto-native financial engineering with distinct risk profiles not found in traditional monetary instruments.

\noindent\textbf{2. Financialization Heightens Complexity and Systemic Interconnectedness.} The notable adoption of ``Derivatives-driven yield'' (25.93\%) and ``External DeFi yield'' (20.37\%) transforms stablecoins from mere payment tokens into actively managed financial products. This evolution inherently breeds complexity and new vectors for systemic risk. These strategies create direct counterparty exposures to derivative providers and critical dependencies on the operational security and economic stability of external DeFi applications (e.g., Aave, Curve). Such deep entanglement implies that failures in these third-party services could trigger cascading solvency issues across multiple stablecoins.

\noindent\textbf{3. Unsustainable Yields Signal Structural Fragility in a Market Segment.} A significant portion of yield-bearing stablecoins (22.22\% combined) rely on inorganic and inherently unsustainable sources: ``Subsidized funds from community'' and ``Secondary token emissions.'' These are not revenue from viable economic activity but are functionally marketing expenses or temporary incentives. This indicates that a segment of the market may be structurally fragile, reliant on bootstrapping growth through mechanisms that are finite, potentially masking underlying economic non-viability until subsidies deplete or emission-based tokens devalue under pressure.

\insight{The integration of yield imposes a ``\textit{dual mandate}'' that reforges stablecoins from passive anchors into complex financial instruments. This transformation is now mainstream: 56.84\% stablecoins offer yield, of which 83.33\% provide returns exceeding the US 10-year Treasury benchmark. Fulfilling this aggressive yield mandate necessitates high-risk financial engineering, evidenced by the significant reliance on derivatives (25.93\%) and external DeFi protocols (20.37\%). This tension between the mandate for stability and the strategies required for high returns introduces a web of market, counter-party, and contagion risks that fundamentally redefines the asset's evolutionary stakes.}

\section{Security Risks}
\label{sec:security_risk}

The systemic importance of stablecoins means their vulnerabilities can trigger cascading failures. This section analyzes the spectrum of security risks afflicting stablecoins, drawing insights from an empirical study of real-world security incidents detailed in Appendix~\ref{app:security_incidents}. Based on 44 significant incidents, we present a statistical breakdown of their root causes, categorizing them into three primary types: Price Fluctuation, Smart Contract Issues, and Peripheral Factors. Each cause is illustrated with a representative case study.
The incidents were selected based on two primary criteria:
\begin{itemize}
    \item Loss exceeding \$100K, collated from reputable sources such as BlockSec Phalcon Explorer~\cite{blocksec}, REKT Database~\cite{rekt}, SlowMist Hacked~\cite{slowmist}, ChainLight Lumos~\cite{chainlight}, and Neptune Mutual database~\cite{neptune}.
    \item Direct relevance to stablecoin failures, excluding incidents where non-stablecoin projects failed due to external stablecoin issues or merely incurred losses denominated in stablecoins.
\end{itemize}

\begin{figure}[!tb]
    \footnotesize
    \begin{subfigure}[t]{0.3\textwidth}
        \begin{tikzpicture}
            \pie[
                % hide number,
                % text=none, 
                radius=2, 
                % explode=0, 
                color={3, 4, red1, red2, red3, green2, green3, green4},
                cloud, text=inside, scale font,
            ]{4.55/, 27.27/, 38.64/, 25/, 6.82/, 9.09/, 4.55/, 2.27/}
        \end{tikzpicture}
    \end{subfigure}
    \hspace{0.1cm}
    \begin{subfigure}[t]{0.17\textwidth}
        \footnotesize
        \begin{tikzpicture}
            \node[draw, fill=1, minimum width=0.2cm, minimum height=0.2cm] (sm1) at (0, 6) {}; 
            \node[right=0.1cm of sm1] {PF - market volatility};
            \node[draw, fill=2, minimum width=0.2cm, minimum height=0.2cm] (sm2) at (0, 5.6) {};
            \node[right=0.1cm of sm2] {PF - price manipulation};

            \node[draw, fill=red1, minimum width=0.2cm, minimum height=0.2cm] (ca1) at (0, 5.2) {}; 
            \node[right=0.1cm of ca1] {SCI - code vulnerability};
            \node[draw, fill=red2, minimum width=0.2cm, minimum height=0.2cm] (ca2) at (0, 4.8) {};
            \node[right=0.1cm of ca2] {SCI - flash loan};
            \node[draw, fill=red3, minimum width=0.2cm, minimum height=0.2cm] (ca3) at (0, 4.4) {};
            \node[right=0.1cm of ca3] {SCI - governance};

            \node[draw, fill=green1, minimum width=0.2cm, minimum height=0.2cm] (smm1) at (0, 4.0) {}; 
            \node[right=0.1cm of smm1] {PeF - rug pull};
            \node[draw, fill=green2, minimum width=0.2cm, minimum height=0.2cm] (smm2) at (0, 3.6) {};
            \node[right=0.1cm of smm2] {PeF - access control};
            \node[draw, fill=green3, minimum width=0.2cm, minimum height=0.2cm] (smm3) at (0, 3.2) {};
            \node[right=0.1cm of smm3] {PeF - impacted fund};
        \end{tikzpicture}
    \end{subfigure}
    \caption{The cause distribution of historical security incidents (where PF refers to price fluctuation, SCI refers to smart contract issue, and PeF refers to peripheral factor).}
    \label{fig:security_incidents}
\end{figure}
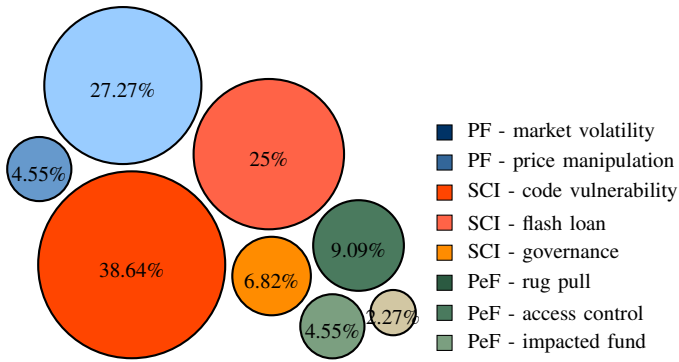

\subsection{Price Fluctuation}

As cryptoassets, stablecoins are inherently exposed to price volatility, a primary concern for both users and attackers. Such fluctuations can be organic, termed market volatility, or maliciously induced, termed price manipulation. These can manifest as gradual drifts, sudden crashes, or even single-transaction shocks, often amplified by tools like flash loans (further discussed in Section~\ref{sec:flashloan}).

\subsubsection{Market Volatility}

Market volatility tests a stablecoin's resilience across three interconnected stress points:
\begin{itemize}
    \item Direct volatility of the stablecoin itself, potentially breaching its peg tolerance.
    \item Devaluation pressure on related protocol tokens (e.g., secondary governance tokens).
    \item Broad downward trends in the wider cryptocurrency market (e.g., BTC).
\end{itemize}
Sustained market volatility can critically undermine a protocol's design, implementation, and public confidence.

\noindent\textbf{Case Study: Terra UST/LUNA.} The algorithmic stablecoin TerraUSD (UST) aimed for a \$1 peg via an arbitrage mechanism with its secondary token, LUNA, where 1 UST was exchangeable for \$1 worth of LUNA~\cite{BRIOLA2023103358}. This design effectively sacrificed LUNA to stabilize UST during de-peg threats. However, significant sell pressure on LUNA triggered a negative feedback loop, i.e., the ``Death Spiral'', leading to UST's collapse in May 2022, despite prior academic discussion of this vulnerability. 

The sheer scale of the Terra/Luna collapse spurred extensive quantitative modeling. However, a significant body of this research concentrated on the broader financial and economic repercussions, such as contagion effects across crypto markets~\cite{BRIOLA2023103358}, the market impact of public disclosures~\cite{ahmed2023course,ahmed2024public}, flight-to-safety dynamics~\cite{anadu2023runs}, and overarching devaluation risks~\cite{eichengreen2023stablecoin,liu2023anatomy}. A critical observation is that many of these analyses, while valuable, often did not deeply simulate the stabilization mechanism's specific failure modes under duress, a crucial aspect for understanding its security vulnerabilities against economic attacks or cascading internal breakdowns.

Nevertheless, several studies offered more granular insights into its failings. For instance, Uhlig~\cite{uhlig2022luna} modeled the crash's progression, highlighting diverse agent behaviors concerning convertibility during the crisis. Kurovskiy et al.~\cite{kurovskiy2023algorithmic} explored why the arbitrage mechanism faltered, pinpointing the detrimental effects of floating redemption fees and critical deficiencies in collateral accessibility and liquidity, which are all key parameters for mechanism resilience. From a more technical simulation perspective, Calandra et al.~\cite{calandra2024making} modeled Terra's transaction dynamics and specific de-peg triggers, providing insights into the operational vulnerabilities that precipitated the collapse.

\subsubsection{Price Manipulation}
\label{sec:price_manipulation}

Price manipulation attacks typically exploit control over a stablecoin's (or its collateral's) reference price sources, which can be:
\begin{itemize}
    \item Centralized sources: price dashboards (e.g., CoinMarketCap~\cite{coinmarketcap_stable}) and CEXs (e.g., Binance~\cite{binance}). These are generally harder to manipulate but can suffer from reporting lags or inter-exchange price inconsistencies.
    \item Decentralized sources: oracles (e.g., Chainlink~\cite{chainlink}) and DEXs (e.g., Uniswap~\cite{uniswap}). Oracles can be vulnerable if their feed sources lack sufficient decentralization or robust validation. DEXs, especially AMM-based ones, can amplify price swings if liquidity pools are shallow, making them targets during periods of high volatility or panic.
\end{itemize}

\noindent\textbf{Case Study: BonqDAO BEUR.} BEUR, an over-collateralized stablecoin pegged to 1 EUR, allowed users to mint BEUR against locked assets. Its vulnerability lay in a permissionless price oracle where the last reported price feed for collateral was considered the spot price. In February 2023, attackers momentarily inflated the price of a collateral asset (WALBT) via this oracle, minted an unearned excess of BEUR, and subsequently halved BEUR's price.

\subsection{Smart Contract Issue}

As blockchain-based applications, stablecoins inherit all common smart contract vulnerabilities, while also presenting unique attack surfaces related to their specific economic logic, governance structures, and interactions facilitated by blockchain features like flash loans.

\subsubsection{Code Vulnerability}

Standard software flaws persist in stablecoin contracts, including reentrancy, insufficient input validation (e.g., Beanstalk, Prisma), and logic errors stemming from ``copy-paste'' practices (e.g., Yearn). The impact of such vulnerabilities is often direct and catastrophic.

\noindent\textbf{Case Study: Cashio CASH.} CASH stablecoins could be minted against Saber LP and Arrow Protocol collateral. A critical flaw in the \texttt{mint} function involved improper validation of the Arrow account and no token matching. In March 2022, an attacker exploited this by using worthless tokens to mint approximately \$53M in CASH, leading to the stablecoin's failure.

\subsubsection{Flash Loan Attack}
\label{sec:flashloan}

Flash loans, which are uncollateralized loans borrowed and repaid within a single atomic transaction~\cite{10.1007/978-3-662-64322-8_1,10272455}, provide attackers with immense temporary capital. While a neutral financial tool, they can be weaponized to exploit vulnerabilities in a protocol's economic logic, price oracle dependencies, or governance mechanisms.

\noindent\textbf{Case Study: Beanstalk BEAN.} In April 2022, an attacker leveraged a flash loan of over \$1 billion (from Aave, Uniswap, SushiSwap) to acquire enough governance tokens to pass malicious Beanstalk Improvement Proposals (BIP18, BIP19). These proposals authorized fund transfers to the attacker. The entire sequence of loan acquisition, voting, proposal execution, and loan repayment occurred within one transaction, resulting in a \$182 million loss and the de-facto failure of BEAN as a stablecoin.

\subsubsection{Governance Attack}
\label{sec:governance_attack}

Blockchain governance aims to enable decentralized decision-making for protocol evolution and safety~\cite{10.1145/3558535.3559794,LIU2023111576}. However, poorly designed, implemented, or managed governance systems can introduce critical vulnerabilities, allowing attackers to manipulate outcomes for malicious profit, with effects that are often hard to reverse.

\noindent\textbf{Case Study: Mochi USDM.} In November 2021, Mochi exploited Curve's governance by using its USDM stablecoin to acquire a large stake in CVX tokens, thereby gaining disproportionate influence over Curve's reward allocations. This allowed Mochi to boost rewards for its own USDM pool, attract significant liquidity, and subsequently drain \$30 million from this pool before abandoning the stablecoin.

\subsection{Peripheral Factor}

Beyond the aforementioned, a range of peripheral yet critical factors contribute to stablecoin security risks.

\subsubsection{Rug Pull}

A rug pull is an exit scam where project founders or developers abruptly abandon the project after attracting capital, leaving investors with worthless tokens~\cite{10515209,287184}. This can occur in both centralized and ostensibly decentralized stablecoin projects, often by exploiting pre-set vulnerabilities, centralized control points, or inadequate access controls over protocol funds or liquidity pools.

\noindent\textbf{Case Study: DEFI100.} A BSC-based synthetic asset index product, DEFI100 executed an apparent exit scam in May 2021. The project's official website briefly displayed a message ``We lied to you, you can't do anything with us'' before being taken offline.

\subsubsection{Access Control}

Compromised access control, often involving private keys that represent ultimate authority over contracts or funds, is a fundamental security threat~\cite{312842}. The security of admin keys, deployer wallets, and multi-signature participants is paramount. Several major stablecoin losses trace back to compromised operational security.

\noindent\textbf{Case Study: Tether.} In November 2017, Tether announced that approximately 31 million USDT were illicitly removed from its treasury wallet due to an external attack compromising access. Notably, despite this significant breach, USDT's market dominance was not fatally impacted in the long term, highlighting complex market reactions to such incidents.

\subsubsection{Impacted Fund}

Stablecoins often rely on, or deposit their reserves/collateral into, other DeFi protocols or custodial solutions to generate yield or manage assets. The security of these external dependencies is crucial; a failure in a third-party application can directly impact the stablecoin's backing or solvency.

\noindent\textbf{Case Study: Angle Protocol.} The Angle Protocol held \$18M USDC in the Euler Protocol. When Euler was hacked for \$197M in March 2023, Angle Protocol's funds on Euler were lost, rendering Angle's stablecoin products under-collateralized. Despite this, Angle Protocol has continued to operate, maintaining a notable market share even today.

\insight{While all DeFi systems evolve, stablecoins undergo a particularly acute evolutionary process, focusing on the core mission of improving peg stability. Designs are thus forged through ongoing trial-and-error, where market adoption, liquidity dynamics, and critical incidents, with most notably technical flaws like code vulnerabilities(38.64\%) and economic stresses from market volatility(27.27\%), act as stringent ``\textit{evolutionary pressures}''. Market feedbacks and security (near)-incidents challenging peg integrity are pivotal. They necessitate crucial adaptations to a stablecoin's peg mechanism for survival or expose fatal flaws causing failure, which then spurs broader re-evaluation of stability models. }

\section{The Stablecoin LEGO Framework}
\label{sec:stablecoin_lego}

This section details the \textbf{Stablecoin LEGO} framework, our proposed methodology for evaluating stablecoin resilience. We will 1) present the architecture and scoring methodology of the framework, and 2) demonstrate its practical utility by applying it to real-world stablecoins, an analysis that yields crucial findings regarding their systemic risks and structural integrity.

\subsection{Motivation}

A robust evaluation of stablecoins demands a framework that addresses their inherent dual nature: they are simultaneously blockchain-based software and nascent monetary instruments. Unlike traditional decentralized applications (DApps), whose primary role is to provide a service or platform, a stablecoin's core objective is to maintain price stability and function as a reliable digital representation of value. Consequently, any rigorous assessment cannot be confined to technical security audits alone; it must also incorporate analytical models from monetary theory and finance to scrutinize the design, resilience, and economic viability of the stability mechanism itself.

\noindent\textbf{Background.} Prior work has initiated the development of stablecoin evaluation frameworks. For instance, Bluechip's SMIDGE framework~\cite{bluechip_2022} assesses stablecoins across broad dimensions but primarily offers high-level safety scores without deep, systematic justification for its scoring logic. Similarly, Moody's Digital Asset Monitor~\cite{moodys_2023} provides sophisticated tracking of on-chain and off-chain events but concentrates on financial market dynamics, giving less weight to the underlying technical security and specific design choices of the protocols.

While these initial frameworks provide a valuable starting point, they generally do not offer a sufficiently granular or compositional approach to risk. To fill this gap, we introduce the \textbf{Stablecoin LEGO} framework, designed for systematic, holistic, and compositional risk evaluation. The name is intentional: complex stablecoins are rarely monolithic. Instead, they are composed of interoperable building blocks (the ``LEGOs''), such as collateral management systems, price oracles, governance modules, and redemption mechanisms. Our framework mirrors this reality. It provides a methodology to first deconstruct a stablecoin into its fundamental components (the internal LEGOs), then evaluate each ``block'' for its specific technical risks and economic assumptions, and finally assess the systemic risks that emerge from their internal composition and external interactions.

\subsection{Methodology}

\begin{figure}[!tb]
    \centering
    \includegraphics[width=\linewidth]{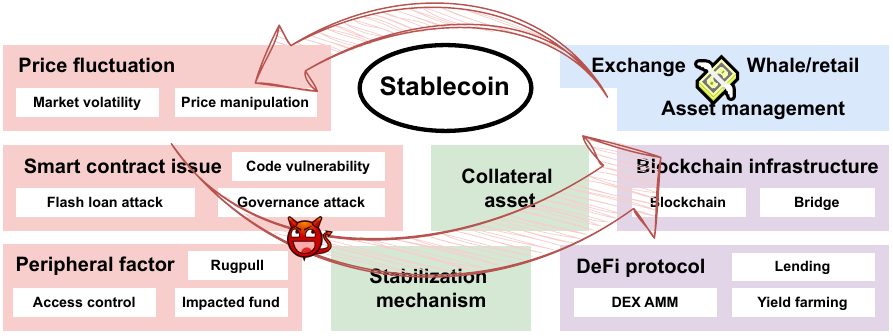}
    \caption{An overview of the Stablecoin LEGO. The left-hand side represents the upstream while the right-hand side represents the downstream.}
    \label{fig:lego}
\end{figure}

Our Stablecoin LEGO framework models a stablecoin as a dynamic system of three interacting core elements, allowing us to analyze how risks propagate through its ecosystem, as illustrated in Fig.~\ref{fig:lego}. These elements are:

\begin{itemize}
    \item \textbf{Upstream Risk Factors (UP(t)):} External market forces and security threat vectors that impose risks upon a stablecoin (e.g., market volatility, smart contract exploits).
    \item \textbf{Stablecoin Intrinsic State (S(t)):} The internal design choices and active mechanisms that determine a stablecoin's inherent resilience and dynamic response to shocks (e.g., collateralization ratio, governance responsiveness).
    \item \textbf{Downstream Ecosystem Composition (DN(t)):} The network of applications, services, and holder concentrations that build upon or rely on the stablecoin, creating pathways for feedback loops.
\end{itemize}

These components interact non-linearly, with feedback loops playing a crucial role (e.g., a major downstream event could trigger a crisis of confidence, impacting the stablecoin's intrinsic state and altering upstream market perception). To capture these complex dynamics, we formalize our framework using a System Dynamics Model, an approach widely adopted for modeling complex socio-economic and financial systems~\cite{martinez2013best, radzicki2020system, systems5020029, PASSARELLA2012570, john2012linking, SENGE1980269}.

\noindent\textbf{Formalization.}
The core of our model is a differential equation that describes the evolution of the stablecoin's state, S(t), over time. For the purpose of this model, S(t) can be represented by a key indicator of its health and scale, such as market capitalization. The state's rate of change is governed by influences from the Upstream (UP(t)), Downstream (DN(t)), and its own internal dynamics:

The core of our model is a differential equation describing the evolution of the stablecoin's intrinsic state, S(t), over time. For the purpose of this model, S(t) can be represented by a key indicator of its health and scale, such as its market capitalization or deviation from peg. The state's rate of change is governed by influences from Upstream risk factors (UP(t)), Downstream ecosystem composition (DN(t)), and its own internal dynamics and resilience mechanisms:
\begin{equation}
    \label{eq:main_revised}
    \frac{d\text{S}(t)}{dt} = \underbrace{\alpha \cdot \text{UP}(t)}_{\text{External shocks}} + \underbrace{\beta \cdot \text{DN}(t)}_{\text{Ecosystem feedback}} + \underbrace{f(\text{S}(t), \text{params})}_{\text{Internal dynamics}}
\end{equation}
where $\alpha$ and $\beta$ are gain coefficients for upstream and downstream inputs, respectively. Our framework then focuses on quantifying the UP(t) and DN(t) components based on empirical data.

\subsubsection{Upstream Risk Factors}

The upstream component, UP(t), quantifies the aggregate external risks facing a stablecoin. We define the sources of these risks, termed Impact Objects, based on common causes of security incidents (detailed in Section~\ref{sec:security_risk}). To measure the severity of each object, we assign it an Impact Degree, a composite score derived from three facets: exposure index (accessibility of the vulnerable component), impact nature (direct/indirect effect), and loss form (e.g., fund loss vs. control loss), as specified in Table~\ref{tab:upstream_impact_degree}. These degrees function as risk weights. The total upstream risk is the weighted sum of all impact objects:
\begin{equation}
    \label{eq:upstream}
    \text{UP}(t) = \sum_{k=1}^n w_k \cdot m_k(t)
\end{equation}
where for each impact object $k$, $m_k(t)$ is its quantified metric (e.g., price deviation, audit status) and $w_k$ is the scalar weight derived from its Impact Degree. Specific metrics and their weighting rationale are in Table~\ref{tab:upstream_metrics}.

\begin{table*}[!tb]
    \centering
    \begin{tabular}{lp{11cm}l}
        \toprule
        Impact degree & Explanation & Notation \\
        \midrule
        \multirow{4}{*}{Exposure index} & Exposure concerning basic blockchain, ecosystem, and trading rules. & $e_1$ \\
         & Exposure concerning the protocol designs of specific applications, yet are publicly accessible, e.g., from blockchain data, documentation, audit reports, and open-sourced code. & \multirow{2}{*}{$e_2$} \\
         & Exposure concerning secret information accessible only within a limited range, e.g., private keys. & $e_3$ \\
        \midrule
        \multirow{3}{*}{Impact nature} & Impact that can indirectly affect the stablecoin. & $i_1$ \\
         & Impact that can directly affect the stablecoin. & $i_2$ \\
         & Impact that can hybrid affect the stablecoin, i.e., directly and indirectly. & $i_3$ \\
        \midrule
        \multirow{3}{*}{Loss form} & The loss is calculated in the form of the number of tokens. & $l_1$ \\
         & The loss is calculated in the form of the price drop. & $l_2$ \\
         & The loss is calculated as a consequence of the loss of control of the stablecoin protocol. & $l_3$ \\
        \bottomrule
    \end{tabular}
    \caption{The definitions of the impact degrees of the upstream risk factors, divided into 3 aspects, each demonstrating 3 levels of impact severity.}
    \label{tab:upstream_impact_degree}
\end{table*}

\begin{table*}[!tb]
    \centering
    \begin{tabular}{llll}
        \toprule
        \multicolumn{2}{l}{Impact object} & Quantification metrics & Impact degree \\
        \midrule
        \multirow{2}{*}{Price fluctuation} & Market volatility & Price standard deviation in the past 5 years & $(e_1,i_3,l_2)$ \\
         & Price manipulation & Regular security auditing & $(e_2,i_3,l_2)$ \\
        \midrule
        \multirow{3}{*}{Smart contract issue} & Code vulnerability & Regular security auditing  & $(e_2,i_1,l_3)$ \\
         & Flash loan attack & Regular security auditing & $(e_2,i_1,l_3)$ \\
         & Governance attack & Regular security auditing and token decentralization & $(e_2,i_1,l_3)$ \\
        \midrule
        \multirow{3}{*}{Peripheral factor} & Rug pull & Regular security auditing and attestation report & $(e_3,i_1,l_3)$ \\
         & Access control & Regular security auditing & $(e_3,i_1,l_3)$ \\
         & Impacted fund & Regular attestation report & $(e_2,i_1,l_1)$ \\
        \bottomrule
    \end{tabular}
    \caption{The quantification metrics of the upstream for the impacted objects, performing in a weighted manner.}
    \label{tab:upstream_metrics}
\end{table*}

\subsubsection{Downstream Ecosystem Composition}

\begin{figure}[!tb]
    \footnotesize
    \begin{tikzpicture}
        \begin{axis}[
            xbar stacked, 
            bar width=5pt, 
            symbolic y coords={USDB, TUSD, FRAX, USDD, PYUSD, USDe, FDUSD, USDS, DAI, USDC, USDT}, 
            ytick=data,  
            xmin=0, 
            xmax=108, 
            width=0.5\textwidth, 
            height=0.3\textwidth, 
            enlarge y limits=0.07,
            legend style={
                at={(0.5, -0.1)}, 
                anchor=north, 
                legend columns=3, 
                column sep=0.1cm
                % row sep=0.1cm 
            },
            legend cell align={left},
            grid=both,
            grid style={dashed,gray!30},
        ]
        \addplot[xbar, fill=green1] coordinates {(0,USDB)(3.9264,TUSD)(0.0165,FRAX)(1.2696,USDD)(17.6987,PYUSD)(8.4785,USDe)(96.7493,FDUSD)(0.0417,USDS)(2.0564,DAI)(14.6538,USDC)(53.2682,USDT)};
        \addplot[xbar, fill=green2] coordinates {(0,USDB)(0.1905,TUSD)(0.2029,FRAX)(0.0054,USDD)(9.9289,PYUSD)(0.5660,USDe)(0.0558,FDUSD)(0.4356,USDS)(3.3415,DAI)(5.5597,USDC)(4.4057,USDT)};
        \addplot[xbar, fill=green3] coordinates {(12.0578,USDB)(0.7820,TUSD)(53.1922,FRAX)(98.0634,USDD)(1.3587,PYUSD)(82.1671,USDe)(0.0207,FDUSD)(70.8696,USDS)(25.0256,DAI)(3.5957,USDC)(0.6603,USDT)}; 
        \addplot[xbar, fill=green4] coordinates {(0.0394,USDB)(0.2251,TUSD)(14.8037,FRAX)(0.1439,USDD)(0.1112,PYUSD)(0.0634,USDe)(0.0169,FDUSD)(0,USDS)(6.3893,DAI)(3.9556,USDC)(6.6690,USDT)};
        \addplot[xbar, fill=green5] coordinates {(82.5321,USDB)(94.2145,TUSD)(31.7717,FRAX)(0.5097,USDD)(70.7349,PYUSD)(8.7129,USDe)(3.1495,FDUSD)(28.6523,USDS)(47.7420,DAI)(47.6325,USDC)(13.7521,USDT)};

        \legend{Exchange, Asset management, DeFi protocol, Blockchain infrastructure, Whale/retail}
        \end{axis}
    \end{tikzpicture}
    \caption{The token distributions of the 11 stablecoins regarding identity and ownership.}
    \label{fig:token_distribution}
\end{figure}
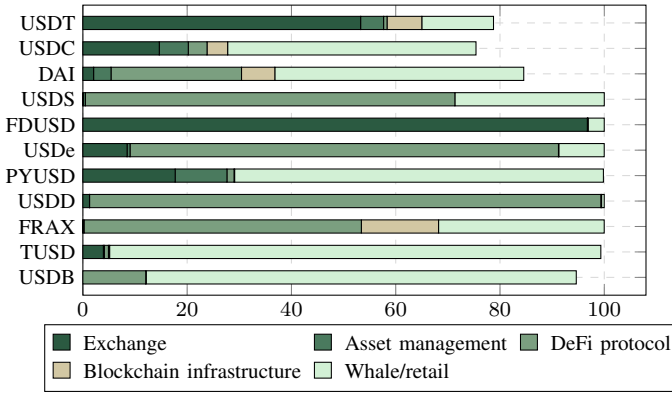

The downstream component, DN(t), captures the concentration and composition of a stablecoin's holders and integrated protocols. This determines the potential ``blast radius'' of a failure and the pathways for contagion: it is critical to understanding how the stablecoin ``LEGO brick'' interlocks with the broader DeFi structure. A stablecoin's distress can trigger chain reactions, and identifying key dependencies is vital.
Our analysis focuses on the top 1000 addresses for each of the 11 stablecoins by market capitalization (see Fig.~\ref{fig:token_distribution} for the evaluated subset). These typically represent over 75\% of the total token supply (while some over 99\%). We identify and categorize these holders (e.g., centralized exchanges, DeFi protocols) using services from Etherscan and BlockSec MetaSuite. The downstream impact is thus represented by a vector of token shares held by each category:
\begin{equation}
    \text{DN}(t) = \mathbf{TokenShare}(t)
\end{equation}

\begin{table*}[!tb]
    \centering
    \begin{tabular}{l|rrrr|rrrrr}
         \toprule
         \multirow{3}{*}{Stablecoin} & \multicolumn{4}{c|}{$\text{UP}(t)$} & \multicolumn{5}{c}{$\text{DN}(t)$} \\
         & \multirow{2}{*}{\makecell[r]{Price\\ fluctuation}} & \multirow{2}{*}{\makecell[r]{Smart contract\\ issue}} & \multirow{2}{*}{\makecell[r]{Peripheral\\ factor}} & \multirow{2}{*}{\makecell[r]{Total}} & \multirow{2}{*}{Exchange} & \multirow{2}{*}{\makecell[r]{Asset\\ management}} & \multirow{2}{*}{\makecell[r]{DeFi\\ protocol}} & \multirow{2}{*}{\makecell[r]{Blockchain\\ infrastructure}} & \multirow{2}{*}{Whale/retail} \\
         & & & & & & & & & \\
         \midrule
         USDT & 2.1583 & 3.7000 & 5.7101 & 12.7117 & \textbf{53.2682} & 4.4057 & 0.6603 & 6.6690 & 13.7521 \\
         USDC & 2.1583 & 3.7000 & 5.6553 & 12.6570 & 14.6538 & 5.5597 & 3.5957 & 3.9556 & 47.6325 \\
         DAI & 1.9833 & 3.4000 & 5.4750 & 11.7492 & 2.0564 & 3.3415 & 25.0256 & 6.3893 & 47.7420 \\
         USDS & 0.0001 & 0.0000 & 2.5000 & \textbf{3.0940} & 0.0417 & 0.4356 & \textbf{70.8696} & 0.0000 & 28.6523 \\
         FDUSD & 2.0417 & 3.5000 & 5.4803 & 12.8872 & \textbf{96.7493} & 0.0558 & 0.0207 & 0.0169 & 3.1495 \\
         USDe & 1.5167 & 2.6000 & 4.3366 & 9.7278 & 8.4785 & 0.5660 & \textbf{82.1671} & 0.0634 & 8.7129 \\
         PYUSD & 4.1583 & 3.7000 & 5.6553 & 14.8439 & 17.6987 & 9.9289 & 1.3587 & 0.1112 & \textbf{70.7349} \\
         USDD & 2.1000 & 3.6000 & 5.6500 & 13.2500 & 1.2696 & 0.0054 & \textbf{98.0634} & 0.1439 & 0.5097 \\
         FRAX & 1.4583 & 2.5000 & 4.6875 & 9.3755 & 0.0165 & 0.2029 & \textbf{53.1922} & 14.8037 & 31.7717 \\
         TUSD & 2.2167 & 3.8000 & 3.3250 & 11.2278 & 3.9264 & 0.1905 & 0.7820 & 0.2251 & \textbf{94.2145} \\
         USDB & 2.3347 & 4.0000 & 6.0000 & 14.1825 & 0.0000 & 0.0000 & 12.0578 & 0.0394 & \textbf{82.5321} \\
         \bottomrule
    \end{tabular}
    \caption{The evaluation results of 11 stablecoins under the Stablecoin LEGO framework.}
    \label{tab:evaluation_results}
\end{table*}

\subsection{Result}

The application of our Stablecoin LEGO framework yields the quantitative risk profiles (initial evaluation on 11 stablecoins, Table~\ref{tab:evaluation_results}). The upstream $UP(t)$ score quantifies inherent protocol and market risks (higher is riskier), while the downstream $DN(t)$ distribution reveals concentration, indicating the nature of systemic risk. This section presents our key findings by deconstructing these results through illustrative case studies and pattern analysis, demonstrating how the framework provides a granular, data-driven view of the stablecoin risk landscape.

\subsubsection{Finding 1: Risk Is Not Monolithic}

Our analysis reveals that stablecoins fall into distinct risk archetypes defined by their downstream composition. The LEGO framework's value lies in its ability to differentiate these risk profiles, as illustrated by the following cases.

\noindent\textbf{Case Study 1: The DeFi-centric archetype (USDD).}
USDD scores a moderately high Upstream risk of 13.25 but, more critically, has a 98.1\% downstream concentration in DeFi protocols. This profile is structurally reminiscent of past failures like Terra UST. While its stabilization mechanism differs, its near-total reliance on host DeFi ecosystems creates enormous contagion risk. The framework flags this clearly: a vulnerability in its stabilization logic (reflected in its Upstream score) would hardly be contained. Its impact would be massively amplified, threatening a cascading failure of the liquidity pools and lending platforms that treat it as a foundational ``LEGO brick.'' The framework thus identifies a critical concern: an extreme dependency on the health and security of a few third-party DeFi protocols.

\noindent\textbf{Case Study 2: The exchange-centric archetype (FDUSD).}
FDUSD presents a different risk profile. Its Upstream score of 12.89 is comparable to USDD's, but its 96.7\% concentration on CEXs shifts the primary threat to custodial and counterparty risk. The framework's downstream analysis highlights that for an FDUSD holder, the security of the stablecoin is less about its on-chain mechanism and more about the solvency, security practices, and regulatory standing of a single corporate entity (e.g., Binance). This recalls historical failures like exchange collapses, where users' assets were frozen or lost. Our framework makes this abstract risk concrete and quantifiable.

\noindent\textbf{Case Study 3: The whale-dominated archetype (TUSD).}
TUSD, with an Upstream risk of 11.23 and a 94.2\% concentration in private whale wallets, exemplifies a third archetype. The primary risk here is the centralization of power and market stability. The framework reveals that the asset's fate rests in the hands of a few large, anonymous holders. This composition makes it structurally vulnerable to a ``bank run'' scenario, where a few entities exiting could collapse market confidence. This insight goes beyond analyzing the protocol's code to assessing the real-world distribution of power over the asset.

\subsubsection{Finding 2: Common Ecosystem-Wide Weaknesses}

Our analysis also reveals that the Peripheral factor (as defined in Table~\ref{tab:upstream_metrics}) is the principal driver of quantifiable risk across almost all stablecoins. Specifically, this category typically accounts for 40-50\% of the total UP(t) risk for major stablecoins like USDT (44.9\%), USDC (44.7\%), and DAI (46.6\%), extending to 80.8\% for USDS (with TUSD as the main exception where ``Smart Contract Issue'' leads).
This empirical finding signifies a potential systemic misalignment between perceived risk focal points (often core contract logic) and the primary sources of measured upstream vulnerability. The persistently high contribution indicates a critical gap in the efficacy or scope of current industry safeguards for these foundational operational and counterparty threats. This strongly suggests a need for re-prioritizing risk management efforts and audit focuses within the stablecoin ecosystem.

\subsubsection{Discussion and Implications}
The findings from our framework have direct implications for key stakeholders in the ecosystem.

\noindent\textbf{For developers and security auditors:} The results advocate for a shift in focus from purely internal code security to a more holistic, compositional risk analysis. For a DeFi-centric coin like USDD, security audits must extend to the host protocols it depends on. For all protocols, the high scores in ``Peripheral factors'' signal an urgent need to bolster defenses around oracles and governance, which are the very ``connective tissue'' between LEGO bricks.

\noindent\textbf{For investors and users:} The LEGO framework transforms the abstract notion of ``risk'' into a tangible choice between different risk models. An investor can now consciously select their exposure: the systemic contagion risk of DeFi-centric assets, the corporate counterparty risk of exchange-centric coins, or the market manipulation risk of whale-dominated tokens. Our framework argues there is no safest stablecoin in absolute terms, only one whose risk profile best aligns with an individual's tolerance.

\noindent\textbf{For Regulators:} The framework provides a data-driven tool for identifying and monitoring sources of systemic risk. The quantifiable downstream concentration metrics (e.g., FDUSD's 96.7\% exchange concentration) can help pinpoint entities that are ``too big to fail'' within the crypto ecosystem, enabling more targeted oversight.

\subsubsection{Limitations and Future Work}

We acknowledge certain limitations. The risk weights in our model are based on an analysis of historical incidents; a formal sensitivity analysis testing the framework's robustness to different weightings would be a valuable next step to further strengthen our findings. Additionally, our model does not currently incorporate qualitative factors like a protocol's age or reputation (the ``Lindy Effect''), which could be an avenue for future research. We advocate for such evaluations to be an enduring commitment for stablecoin integrity. To this end, we will continue to advance the Stablecoin LEGO framework and periodically update its results, contributing to the secure and sustainable growth of the ecosystem.

\section{Conclusion}

This SoK facilitates the understanding of stablecoins by analyzing 157 research studies, 95 active stablecoins, and 44 major security incidents. We establish that stability is not an inherent property but a fragile equilibrium governed by risk specialization, i.e., design trade-offs that concentrate unmitigated risks. This tension is now systemically exacerbated by a ``dual mandate'' for both stability and high-risk yield. To assess these dynamics, we introduce the Stablecoin LEGO framework, a quantitative methodology for risk evaluation. Ultimately, this work provides a rigorous foundation for building, analyzing, and regulating stablecoins.

\bibliographystyle{IEEEtran}
\bibliography{references}



% \end{thebibliography}

\appendix
% % \appendix
% \section*{Aaa} 

\begin{table}[!tb]
    \centering
    \footnotesize
    \begin{tabular}{lll}
        \toprule
        Media & URL & Visits \\
        \midrule
        CoinTelegraph & https://cointelegraph.com/ & 235.6M \\ % 9.580M
        CoinDesk & https://www.coindesk.com/ & 232.4M \\ % 5.414M
        BeInCrypto & https://beincrypto.com/ & 124.0M \\
        Cryptonews & https://cryptonews.com/ & 64.3M \\
        Decrypt & https://decrypt.co/ & 62.0M \\ % 1.836M
        CoinGape & https://coingape.com/ & 46.1M \\
        Crypto News & https://crypto.news/ & 42.7M \\
        Bitcoin.com News & https://news.bitcoin.com/ & 41.2M \\
        The Crypto Basic & https://thecryptobasic.com/ & 39.0M \\
        U.Today & https://u.today/ & 35.5M \\
        The Block & https://www.theblock.co/ & 29.1M \\ % 907,194
        Bitcoinist & https://bitcoinist.com/ & 23.5M \\
        CryptoSlate & https://cryptoslate.com/ & 20.4M \\ % 793,983 
        CryptoPotato & https://cryptopotato.com/ & 18.3M \\
        Blockworks & https://blockworks.co/ & 17.3M \\
        BlockBeats & https://www.theblockbeats.info/ & 13.8M \\ % 836,280
        Bitcoin Magazine & https://bitcoinmagazine.com/ & 13.2M \\ % 576,120
        NewsBTC & https://www.newsbtc.com/ & 12.9M \\
        Foresight News & https://foresightnews.pro/ & 12.6M \\ % 736,840
        Crypto Daily & https://cryptodaily.co.uk/ & 11.3M \\
        % Crypto Briefing & https://cryptobriefing.com/ & 3.4M \\
        % PANews & https://www.panewslab.com/ & 7.3M \\ % 625,757
        % Odaily & https://www.odaily.news/ & 8.3M \\ % 594,638
        % TechFlow & https://www.techflowpost.com/ & 1.9M \\ % 418,797
        % Forbes Digital Assets & https://www.forbes.com/digital-assets/ & 5.6M \\
        % The Crypto Times & https://www.cryptotimes.io/ & 9.7M \\
        % The Tokenist & https://tokenist.com/ & 1.6M \\
        % Coin Bureau & https://coinbureau.com/ & 6.4M \\
        % The Defiant & https://thedefiant.io/ & 3.2M \\
        % Bankless & https://www.bankless.com/ & 5.4M \\
        % Bloomberg Crypto & https://www.bloomberg.com/crypto & 471.4K \\
        % CoinGecko Blog & https://blog.coingecko.com/ & 63.1K \\
        % CoinCodex News & https://coincodex.com/news/ & 29.7K \\
        \bottomrule
    \end{tabular}
    \caption{The visits count of Web3 medias.}
    \label{tab:web3_media}
\end{table}

\begin{table}[!tb]
    \centering
    \footnotesize
    \begin{tabular}{llll}
         \toprule
         No. & \makecell[l]{Country/\\Organization} & Department & \# \\
         \midrule
         1 & Canada & Bank of Canada & 3 \\
         2 & USA & The Federal Reserve & 9 \\
         3 & UK & Bank of England & 2 \\
         4 & France & Banque de France & 2 \\
         5 & Germany & Deutsche Bundesbank & 6 \\
         6 & Italy & Banca d'Italia & 0 \\
         7 & Japan & Bank of Japan & 0 \\
         8 & Brazil & Banco Central do Brasil & 2 \\
         9 & Russia & Bank of Russia & 3 \\
         10 & India & Reserve Bank of India & 2 \\
         11 & \makecell[l]{China} & \makecell[l]{People's Bank of China, \\Hong Kong Monetary Authority} & 4 \\
         12 & South Africa & South African Reserve Bank & 2 \\
         13 & Mexico & Banco de México & 1 \\
         14 & Argentina & Central Bank of Argentina & $*$ \\
         15 & Türkiye & Türkiye Cumhuriyet Merkez Bankası & 0 \\
         16 & South Korea & Bank of Korea & 3 \\
         17 & Indonesia & Bank Indonesia & 1 \\
         18 & Australia & Reserve Bank of Australia & 13 \\
         19 & Saudi Arabia & Saudi Central Bank & 0 \\
         20 & European Union & European Central Bank & 5 \\
         21 & African Union & $**$ & $**$ \\
         22 & \multicolumn{2}{l}{International Monetary Fund (IMF)} & 6 \\
         23 & \multicolumn{2}{l}{World Bank (WB)} & 2 \\
         24 & \multicolumn{2}{l}{Bank for International Settlements (BIS)} & 8 \\
         25 & \multicolumn{2}{l}{Financial Stability Board (FSB)} & 10 \\
         26 & \multicolumn{2}{l}{Financial Action Task Force (FATF)} & 3 \\
         27 & \multicolumn{2}{l}{Group of Seven (G7)} & 1 \\
          & & Total & 81 \\
         \bottomrule
    \end{tabular}
    \caption{The numbers of the stablecoin-related reports from G20 members and relevant international financial organizations since 2019. $^*$ No English version for the websites and publication. $^{**}$ The central bank related institutions have not yet established. Note that the sum exceeds 81 due to several inter-institution cooperation works. }
    \label{tab:my_label}
\end{table}

\subsection{Stablecoin Definition}
\label{app:definition}

\subsubsection{Prior Research Dataset}
The stablecoin definitions are from academic papers (Table~\ref{tab:def_aca}), governmental reports (Table~\ref{tab:def_ins1},~\ref{tab:def_ins2}), and industry reports (Table~\ref{tab:def_ind}).

\begin{table*}[!tb]
    \centering
    \scriptsize
    \makebox[\textwidth][c]{
    \begin{tabular}{cp{5cm}llp{5cm}p{3cm}}
        \toprule
        No. & Research source & Year & Blockchain & Pegged asset & Stability \\
        \midrule
        1 & Yujin Potter et al. (ICBC)~\cite{10634419} & 2024 & Unspecified & Unspecified & Minimize price fluctuations \\
        2 & Yizhou Cao et al. (SSRN)~\cite{designing_stablecoins} & 2024 & Public blockchain & Stable financial assets & Pegged \\
        3 & Cheick Tidiane Ba et al. (arXiv)~\cite{ba2024investigatingshockingeventsethereum} & 2024 & Unspecified & Unspecified & Pegging \\
        4 & Kun Duan et al. (Finance Res. Lett.)~\cite{DUAN2023103573} & 2023 & Unspecified & Fiat currencies or assets that are relatively stable & Maintain a peg \\
        5 & Richard K. Lyons et al. (J. Int. Money Finance)~\cite{LYONS2023102777} & 2023 & Unspecified & National currency & Lower volatility \\
        6 & Ingo Fiedler et al. (Emerald)~\cite{fiedler2023stablecoins} & 2023 & Unspecified & Fiat currencies like the dollar or physical assets like gold & Pegged \\
        7 & Lennart Ante et al. (FinTech)~\cite{fintech2010003} & 2023 & Unspecified & Other assets, most often the U.S. dollar but also other fiat currencies or physical assets, such as gold & Peg their value \\
        8 & Yiming Ma et al. (SSRN)~\cite{ma2023stablecoin} & 2023 & Blockchain & \$1 (fiat) & Stable \\
        9 & Yongqi Guan et al. (SOUPS Poster)~\cite{guan2023examining} & 2023 & Unspecified & A specific asset & Anchored (fixed value) \\
        10 & Bruce Mizrach (arXiv)~\cite{mizrach2023stablecoinssurvivorshiptransactionscosts} & 2023 & Distributed ledger & Fiat assets and other stores of value & Maintain price stability \\
        11 & Blanka Łęt et al. (Technol. Forecast. Soc. Change)~\cite{LET2023122318} & 2023 & Distributed ledger & An underlying asset, e.g., the US dollar, precious metals & Pegged \\
        12 & Anneke Kosse et al. (BIS)~\cite{kosse2023will} & 2023 & Unspecified & A specified peg & Maintain a stable value \\
        13 & Christoph Bertsch (Riksbank)~\cite{bertsch2023stablecoins} & 2023 & Unspecified & Unspecified & Stable \\
        14 & Christian Catalini et al. (Annu. Rev. Financ. Econ.)~\cite{catalini2022some} & 2022 & Unspecified & A reference asset (typically the US dollar) & Trade at par \\
        15 & Ye Li et al. (SSRN)~\cite{li2022money} & 2022 & Unspecified & Fiat currency & Maintain a stable price \\
        16 & Binh Nguyen Thanh et al. (J. Ind. Bus. Econ.)~\cite{thanh2023stabilities} & 2022 & Unspecified & Another asset & Have stable value \\
        17 & Ariah Klages-Mundt et al. (Math. Financ.)~\cite{klages2022while} & 2022 & Unspecified & Unspecified & Stabilize price/purchasing power \\
        18 & Lin William Cong et al. (JFE)~\cite{CONG2022972} & 2022 & Unspecified & Unspecified & Unspecified \\
        19 & Adrien d'Avernas et al. (SSRN)~\cite{d2022can} & 2022 & Unspecified & An official currency & Maintain a peg \\
        20 & Martin M. Bojaj et al. (Econ. Model.)~\cite{BOJAJ2022105792} & 2022 & Blockchain & Various currencies and commodities & One-to-one peg \\
        21 & Jamie Morgan (RIBAF)~\cite{MORGAN2022101716} & 2022 & Unspecified & A reference asset (typically a fiat currency such as the US\$) & Stabilised \\
        22 & Gary B. Gorton et al. (NBER)~\cite{gorton2022leverage} & 2022 & Unspecified & Fiat currency & Maintain a constant dollar price \\
        23 & Harald Uhlig (NBER)~\cite{uhlig2022luna} & 2022 & Unspecified & Unspecified & Unspecified \\
        24 & Garth Baughman et al. (Fed)~\cite{feds} & 2022 & Unspecified & Real-world asset & Maintain its peg \\
        25 & Gordon Y. Liao et al. (Fed)~\cite{feds2} & 2022 & Distributed ledger & An external reference, typically the U.S. dollar & Peg their value \\
        26 & Parma Bains et al. (IMF)~\cite{bains2022regulating} & 2022 & Unspecified & Specified asset(s) & Maintain a stable value \\
        27 & Wilko Bolt et al. (DNB)~\cite{bolt2022getting} & 2022 & Unspecified & Fiat currency(ies), commodity(ies), cryptoasset(s), or a combination & Maintain a stable value \\
        28 & Lennart Ante et al. (Finance Res. Lett.)~\cite{ANTE2021101867} & 2021 & Unspecified & Less volatile assets or currencies & Pegged \\
        29 & Dirk G. Baur et al. (Finance Res. Lett.)~\cite{BAUR2021101431} & 2021 & Unspecified & Other (relatively) stable assets such as gold or the US dollar & Pegged \\
        30 & Lai T. Hoang et al. (Eur. J. Finance)~\cite{Hoang01112024} & 2021 & Unspecified & Currencies or assets that are (relatively) stable such as the US dollar & Pegged \\
        31 & Lennart Ante et al. (TFSC)~\cite{ANTE2021120851} & 2021 & Public blockchain & Non-volatile values, most commonly a fiat currency & Peg \\
        32 & Ariah Klages-Mundt et al. (CES)~\cite{Klages2021In} & 2021 & Public blockchain & Unspecified & Stabilize the purchasing power \\
        33 & Klaudia Jarno et al. (J. Risk Financial Manag.)~\cite{jrfm14020042} & 2021 & Unspecified & Unspecified & Minimize fluctuations \\
        34 & Christian Catalini et al. (SSRN)~\cite{catalini2021economic} & 2021 & Unspecified & A reference asset, typically the U.S. Dollar & Maintain stability \\
        35 & Cangshu Li et al. (CEJ)~\cite{doi:10.1080/17538963.2021.1872167} & 2021 & Public blockchain & Legal tender or other assets & Relatively stable price \\
        36 & Ingolf G. A. Pernice (FC Workshop)~\cite{pernice2021stablecoin} & 2021 & Unspecified & Unspecified & Close to the price \\
        37 & Wenqi Zhao et al. (FC Workshop)~\cite{10.1007/978-3-662-63958-0_8} & 2021 & Unspecified & External assets & Minimize the volatility \\
        38 & Yujin Kwon et al. (SSRN)~\cite{potter2021trilemma} & 2021 & Unspecified & Unspecified & Provide a stable value \\
        39 & Amani Moin et al. (FC)~\cite{10.1007/978-3-030-51280-4_11} & 2020 & Unspecified & Some reference point, such as USD & Stable \\
        40 & Ariah Klages-Mundt et al. (AFT)~\cite{10.1145/3419614.3423261} & 2020 & Unspecified & Unspecified & Stabilize price\&purchasing power \\
        41 & Jess Cheng (BBLJ)~\cite{cheng2020build} & 2020 & Distributed ledger & A reference asset or basket of assets & Stabilize the price \\
        42 & Makiko Mita et al. (IJSKM)~\cite{ito2020stablecoin} & 2020 & Unspecified & Stable assets or major fiat currencies & Peg \\
        43 & Clemens Jeger et al. (BCCA)~\cite{9274450} & 2020 & Unspecified & Fiat currencies, gold or even another cryptocurrency & Maintain a stable value \\
        44 & Alexander Lipton et al. (arXiv)~\cite{lipton2020tetherlibrastablecoinsdigital} & 2020 & Unspecified & A target quote currency & Low price volatility \\
        45 & Alexander Lipton et al. (Building the New Economy)~\cite{lipton202011} & 2020 & Unspecified & A target quote currency & Low price volatility \\
        46 & Gang-Jin Wang et al. (RIBAF)~\cite{WANG2020101225} & 2020 & Unspecified & A fiat currency (e.g., USD and CNY) or a commodity (e.g., precious metals such as gold and silver) & Low-volatility \\
        47 & Mario Bellia et al. (SSRN)~\cite{bellia2020makes} & 2020 & Unspecified & Unspecified & Unspecified \\
        48 & Fiona van Echelpoel et al. (ECB)~\cite{van2020stablecoins} & 2020 & Unspecified & Currency(ies) & Minimise fluctuations \\
        49 & Jon Frost et al. (DNB)~\cite{frost2020early} & 2020 & Unspecified & Assets or fiat currencies & Maintain a stable value \\
        50 & Douglas W. Arner et al. (BIS)~\cite{arner2020stablecoins} & 2020 & Unspecified & Fiat currencies or other assets & Tied \\
        51 & Makiko Mita et al. (IIAI-AAI)~\cite{8992735} & 2019 & Blockchain & Another currency & Lower volatility \\
        52 & E. L. Sidorenko (ISCDTE)~\cite{10.1007/978-3-030-27015-5_75} & 2019 & Unspecified & Underlying asset (national currency, gold, oil, etc.) & Low volatility \\
        53 & Aleksander Berentsen et al. (VoxEU.org)~\cite{berentsen2019stablecoins} & 2019 & Unspecified & Unspecified & Minimise price volatility \\
        54 & Amani Moin et al. (arXiv)~\cite{moin2019classificationframeworkstablecoindesigns} & 2019 & Unspecified & Some reference point & Stable \\
        55 & Barry Eichengreen (NBER)~\cite{eichengreen2019commodity} & 2019 & Unspecified & Official numeraire & Maintain a peg \\
        56 & Dirk Bullmann et al. (ECB)~\cite{bullmann2019search} & 2019 & Unspecified & Unspecified & Minimise fluctuations \\
        \bottomrule
    \end{tabular}
    }
    \captionsetup{position=below}
    \caption{Stablecoin definitions from academic papers. Note that all descriptions in the last three columns are directly \emph{quoted} from the original text, except for the ``Unspecified''s.}
    \label{tab:def_aca}
\end{table*}

\begin{table*}[!tb]
    \centering
    % \tiny
    \scriptsize
    \makebox[\textwidth][c]{
    \begin{tabular}{cp{4.5cm}lp{2cm}p{5.5cm}p{3cm}}
        \toprule
        No. & Research source & Year & Blockchain & Pegged asset & Stability \\
        \midrule
        1 & The Federal Reserve~\cite{us_2024_1} & 2024 & Unspecified & National currency or another reference asset & Maintain a stable value \\
        2 & The Federal Reserve~\cite{us_2024_2} & 2024 & Unspecified & National currency or another reference asset & Maintain a stable value \\
        3 & Banque de France~\cite{fr_2024} & 2024 & Cryptographic tech. & A benchmark asset (gold, the euro, the dollar, a group of currencies, etc.) & More stable value \\
        4 & Bank of Russia~\cite{ru_2024} & 2024 & Unspecified & Fiat currency and other assets (gold, other commodities, cryptocurrencies, etc.) or a basket thereof & Pegged \\
        5 & Reserve Bank of India~\cite{in_2024} & 2024 & Unspecified & A numeraire like fiat currency or gold & Maintain a fixed face value \\
        6 & Hong Kong Monetary Authority~\cite{hk_2024_1} & 2024 & Blockchain & Certain asset(s), typically fiat currencies & Maintain a stable value \\
        7 & Financial Services and the Treasury Bureau, and Hong Kong Monetary Authority~\cite{hk_2024_2} & 2024 & Decentralised distributed ledger or similar tech. & Fiat currencies and other types of assets & Unspecified \\
        8 & Bank of Korea~\cite{kr_2024} & 2024 & Unspecified & Reserve assets, such as a fiat currency or a commodity & Maintain a stable value \\
        9 & International Monetary Fund~\cite{imf_2024} & 2024 & Unspecified & Specific currencies, such as the U.S. dollar & Pegged \\
        10 & The Federal Reserve~\cite{us_2023_1} & 2023 & Unspecified & National currency or another reference asset & Maintain a stable value \\
        11 & The Federal Reserve~\cite{us_2023_2} & 2023 & Unspecified & National currency or another reference asset & Maintain a stable value \\
        12 & Bank of England~\cite{uk_2023} & 2023 & Unspecified & Fiat currency & Maintain a stable value \\
        13 & Banque de France~\cite{fr_2023} & 2023 & Public blockchain & Fiat currency & Maintain a stable value \\
        14 & Deutsche Bundesbank~\cite{de_2023} & 2023 & Unspecified & Government currencies, asset backing, and crypto tokens & Stable \\
        15 & Bank of Canada~\cite{canada_2023} & 2023 & Unspecified & Fiat currency & Unspecified \\
        16 & Bank of Russia~\cite{ru_2023} & 2023 & Unspecified & Another asset (fiat currency, precious metals, etc.) or a basket of various assets & Maintain a stable value \\
        17 & Banco Central do Brasil~\cite{br_2023} & 2023 & Unspecified & A predefined asset or an asset basket & Peg \\
        18 & South African Reserve Bank~\cite{za_2023} & 2023 & Unspecified & A specified asset, or a pool or basket of assets & Maintain a stable value \\
        19 & Hong Kong Monetary Authority~\cite{hk_2023} & 2023 & Unspecified & A specified asset, or a pool or basket of assets & Maintain a stable value \\
        20 & Bank of Korea~\cite{kr_2023_1} & 2023 & Unspecified & Reserve assets, including currencies and commodities & Achieve price stability \\
        % 21 & Bank of Korea~\cite{kr_2023_2} & 2023 & Unspecified & A specified asset, or a pool or basket of assets & Maintain a stable value \\
        21 & Reserve Bank of Australia~\cite{au_2023} & 2023 & Unspecified & A specified unit of account or store of value, such as a national currency or commodity & Maintain a stable value \\
        22 & Bank for International Settlements~\cite{bis_2023_1} & 2023 & Unspecified & A specified asset, or a pool or basket of assets & Maintain a stable value \\
        23 & Bank for International Settlements and Hong Kong Monetary Authority~\cite{bis_2023_2} & 2023 & Unspecified & A specified asset (typically USD), or a pool or basket of assets & Maintain a stable value \\
        24 & Bank for International Settlements~\cite{bis_2023_3} & 2023 & Blockchain & A specified asset, or a pool or basket of assets & Maintain a stable value \\
        25 & Financial Stability Board~\cite{fsb_2023_1} & 2023 & Unspecified & A specified asset, or a pool or basket of assets & Maintain a stable value \\
        26 & Financial Stability Board and International Monetary Fund~\cite{fsb_2023_2} & 2023 & Unspecified & A specified asset, or a pool or basket of assets & Maintain a stable value \\
        27 & Financial Stability Board~\cite{fsb_2023_3} & 2023 & Unspecified & A specified asset, or a pool or basket of assets & Maintain a stable value \\
        28 & Financial Stability Board~\cite{fsb_2023_4} & 2023 & Unspecified & A specified asset, or a pool or basket of assets & Maintain a stable value \\
        29 & The Federal Reserve~\cite{us_2022_1} & 2022 & Unspecified & One or more assets & Peg \\
        30 & The Federal Reserve~\cite{us_2022_2} & 2022 & Unspecified & National currency or another reference asset & Maintain a stable value \\
        31 & The Federal Reserve~\cite{us_2022_3} & 2022 & Unspecified & National currency or another reference asset & Maintain a stable value \\
        32 & European Central Bank~\cite{adachi2022stablecoins} & 2022 & Unspecified & Official currency(ies) or other assets & Maintain a stable value \\
        33 & Bank of Canada~\cite{canada_2022} & 2022 & Unspecified & National currency in most cases & Less volatile than other cryptoassets \\
        34 & Banco Central do Brasil~\cite{br_2022} & 2022 & Unspecified & One or more assets (such as sovereign currencies or another asset that is not traded in a cryptocurrency trading environment) & Linked \\
        35 & Bank of Russia~\cite{ru_2022} & 2022 & Unspecified & Various assets (fiat currency, precious metals and others) or a basket of various assets & Maintain a stable value \\
        36 & Reserve Bank of India~\cite{in_2022} & 2022 & Unspecified & A specified asset (typically US dollars), or a pool or basket of assets & Maintain a stable value \\
        37 & Bank of Korea~\cite{kr_2022} & 2022 & Unspecified & A specific asset (usually a fiat currency) & Stabilize the value \\
        38 & Bank Indonesia~\cite{id_2022} & 2022 & Unspecified & A commodity or currency & Relatively stable \\
        39 & Reserve Bank of Australia~\cite{au_2022_1} & 2022 & Unspecified & A specified unit of account or store of value & Maintain a stable value \\
        40 & Reserve Bank of Australia~\cite{au_2022_2} & 2022 & Unspecified & Fiat currencies (particularly the US dollar) or other assets (such as gold) & Maintain a stable value \\
        41 & Reserve Bank of Australia~\cite{au_2022_3} & 2022 & Unspecified & One or more fiat currencies or assets (e.g. the US dollar or gold) & Maintain a stable value \\
        42 & Reserve Bank of Australia~\cite{au_2022_4} & 2022 & Unspecified & Another asset or a basket of assets – commonly a fiat currency (e.g. the US dollar) or a common store of value (e.g. gold) & Minimise price volatility \\
        43 & European Central Bank~\cite{eu_2022_1} & 2022 & Unspecified & One or several official currencies or other assets (including crypto-assets) & Maintain a stable value \\
        44 & European Central Bank~\cite{eu_2022_2} & 2022 & Unspecified & One or several currencies or other assets (including crypto-assets) & Maintain a stable value \\
        45 & European Central Bank~\cite{eu_2022_3} & 2022 & Unspecified & Typically a single fiat currency (or a basket of fiat currencies) & Minimise price volatility \\
        46 & International Monetary Fund~\cite{imf_2022_1} & 2022 & Unspecified & Usually a fiat currency & Maintain stable value \\
        47 & International Monetary Fund~\cite{imf_2022_2} & 2022 & Unspecified & A stable reference asset & Pegged \\
        48 & Bank for International Settlements~\cite{bis_2022_1} & 2022 & Unspecified & A specified asset, or a pool or basket of assets & Maintain a stable value \\
        49 & Bank for International Settlements~\cite{bis_2022_2} & 2022 & Unspecified & A specified asset, or a pool or basket of assets & Maintain a stable value \\
        50 & Financial Stability Board~\cite{fsb_2022_1} & 2022 & Unspecified & A specified asset, or a pool or basket of assets & Maintain a stable value \\
        51 & Financial Stability Board~\cite{fsb_2022_2} & 2022 & Unspecified & A specified asset, or a pool or basket of assets & Maintain a stable value \\
        52 & Financial Stability Board~\cite{fsb_2022_3} & 2022 & Unspecified & A specified asset (typically US dollars), or basket of assets & Maintain a stable value \\
        53 & The Federal Reserve~\cite{us_2021_1} & 2021 & Distributed ledger & National currency or other reference asset or assets & Maintain a stable value \\
        54 & Bank of England~\cite{uk_2021} & 2021 & Unspecified & Government-sponsored or `fiat' currencies & Peg \\
        55 & Deutsche Bundesbank~\cite{de_2021_1} & 2021 & Unspecified & A reference value & Stabilised \\
        56 & Deutsche Bundesbank~\cite{de_2021_2} & 2021 & Distributed ledger & A reference value & Be as stable in value as possible \\
        57 & Deutsche Bundesbank~\cite{de_2021_3} & 2021 & Distributed ledger & Another unit of value & Minimise major fluctuations \\
        \bottomrule
    \end{tabular}
    }
    \captionsetup{position=below}
    \caption{Stablecoin definitions from governmental institution reports, including government agencies and international organizations. Note that all descriptions in the last three columns are directly \emph{quoted} from the original text, except for the ``Unspecified''s.}
    \label{tab:def_ins1}
\end{table*}

\begin{table*}[!tb]
    \centering
    % \tiny
    \scriptsize
    \makebox[\textwidth][c]{
    \begin{tabular}{cp{4.5cm}lp{2cm}p{5.5cm}p{3cm}}
        \toprule
        No. & Research source & Year & Blockchain & Pegged asset & Stability \\
        \midrule
        58 & Bank of Canada~\cite{canada_2021} & 2021 & Unspecified & A basket of assets & Less volatile \\
        59 & South African Reserve Bank~\cite{za_2021} & 2021 & Unspecified & Another asset (typically a unit of currency or commodity) or a basket of assets & Maintain a stable value \\
        60 & Banco de México~\cite{mx_2021} & 2021 & Distributed registry & Unspecified & Minimize fluctuation \\
        61 & Reserve Bank of Australia~\cite{au_2021_1} & 2021 & Unspecified & A specified asset or pool of assets & Maintain a stable value \\
        62 & Reserve Bank of Australia~\cite{au_2021_2} & 2021 & Unspecified & Unspecified & Maintain a stable value \\
        63 & Reserve Bank of Australia~\cite{au_2021_3} & 2021 & Unspecified & One or more currencies or assets & Maintain a stable value \\
        64 & Bank for International Settlements, International Monetary Fund, and World Bank~\cite{bis_imf_wb_2021} & 2021 & Unspecified & A specified asset, or a pool or basket of assets & Maintain a stable value \\
        65 & Financial Action Task Force~\cite{fatf_2021} & 2021 & Unspecified & Some reference asset or assets & Maintain a stable value \\
        66 & Reserve Bank of Australia~\cite{au_2020_1} & 2020 & Unspecified & Another asset, typically a unit of currency or a commodity & Maintain a stable value \\
        67 & Reserve Bank of Australia~\cite{au_2020_2} & 2020 & Unspecified & A widely used unit of account (such as the US dollar) or a common store of value (such as gold) & Minimise price volatility \\
        68 & Bank for International Settlements and World Bank~\cite{wb_2020} & 2020 & Unspecified & Currency/ies & Minimise fluctuations \\
        69 & Financial Stability Board~\cite{fsb_2020_1} & 2020 & Unspecified & A specified asset, or a pool or basket of assets & Maintain a stable value \\
        70 & Financial Stability Board~\cite{fsb_2020_2} & 2020 & Unspecified & A specified asset, or a pool or basket of assets & Maintain a stable value \\
        71 & Financial Action Task Force~\cite{fatf_2020_1} & 2020 & Unspecified & Reference assets & Maintain a stable value \\
        72 & Financial Action Task Force~\cite{fatf_2020_2} & 2020 & Unspecified & Some reference asset or assets & Maintain a stable value \\
        73 & The Federal Reserve~\cite{us_2019_1} & 2019 & Unspecified & An underlying asset or basket of assets & Tied \\
        74 & Deutsche Bundesbank~\cite{de_2019_1} & 2019 & Unspecified & Unspecified & Maintain a stable value \\
        75 & Deutsche Bundesbank~\cite{de_2019_2} & 2019 & Unspecified & Often an existing currency (or basket of currencies) & Have a stable value \\
        76 & Reserve Bank of Australia~\cite{au_2019_1} & 2019 & Unspecified & Unit of account (often the US dollar) or a common store of value (such as gold) & Minimise price volatility \\
        77 & Reserve Bank of Australia~\cite{au_2019_2} & 2019 & Unspecified & Another asset, typically a unit of currency or a commodity & Maintain a stable value \\
        78 & Reserve Bank of Australia~\cite{au_2019_3} & 2019 & Unspecified & A reference asset (such as a sovereign currency or gold) or a basket of assets & Minimise price volatility \\
        79 & European Central Bank~\cite{eu_2019} & 2019 & Unspecified & Currency(ies), securities\&commodities, crypto-assets, and future expectations & Minimise price fluctuations \\
        80 & Financial Stability Board~\cite{fsb_2019} & 2019 & Unspecified & Another asset (typically a unit of currency or commodity) or a basket of assets & Maintain a stable value \\
        81 & Group of Seven, International Monetary Fund, and Bank for International Settlements~\cite{g7_2019} & 2019 & Distributed ledger & Fiat currencies & Achieve stable value \\ 
        \bottomrule
    \end{tabular}
    }
    \captionsetup{position=below}
    \caption{(Cont'd) Stablecoin definitions from governmental institution reports, including government agencies and international organizations. Note that all descriptions in the last three columns are directly \emph{quoted} from the original text, except for the ``Unspecified''s.}
    \label{tab:def_ins2}
\end{table*}

\begin{table*}[!tb]
    \centering
    % \tiny
    \scriptsize
    \makebox[\textwidth][c]{
    \begin{tabular}{cp{4cm}llp{5.5cm}p{3.5cm}}
        \toprule
        No. & Research source & Year & Blockchain & Pegged asset & Stability \\
        \midrule
        1 & Cointelegraph~\cite{cointelegraph_2024} & 2024 & Blockchain & Fiat currencies & Offer price stability \\ 
        2 & IDA and Quinlan\&Associates~\cite{ida_2024} & 2024 & Distributed ledger & Fiat currency values & Ensure close alignment \\
        3 & Visa~\cite{visa_2024} & 2024 & Blockchain & Unspecified & Maintain a stable value \\
        4 & BeInCrypto~\cite{beincrypto_2024} & 2024 & Unspecified & Another asset, such as gold, fiat currency, or another cryptocurrency & Maintain a set (near-constant) value \\
        5 & Standard Chartered and Zodia Markets~\cite{chartered_2024} & 2024 & Unspecified & A national currency or other reference rate & Maintain a stable value \\ 
        6 & CoinDesk~\cite{coindesk_2024} & 2024 & Unspecified & Another asset class, such as a fiat currency or gold & Keep a stable, steady value \\
        7 & Castle Island Ventures and Brevan Howard Digital~\cite{castle_2024} & 2024 & Public blockchain & Fiat currency & Unspecified \\
        8 & Chainalysis\cite{chainalysis_2024_1} & 2024 & Unspecified & Unspecified & Unspecified \\
        9 & Chainalysis\cite{chainalysis_2024_2} & 2024 & Unspecified & Typically U.S. dollar & Pegged \\
        10 & CCData~\cite{ccdata_2024} & 2024 & Unspecified & Another currency, commodity, or financial instrument & Pegged \\
        11 & Stablecoin Standard~\cite{stablecoin_standard_2024} & 2024 & Blockchain & Fiat or e-money & Unspecified \\
        12 & PwC and Stellar Development Foundation~\cite{pwc_2023_1} & 2023 & Unspecified & Fiat currencies, commodities or other crypto assets & Price stability \\
        13 & PwC~\cite{pwc_2023_2} & 2023 & Unspecified & Unspecified & Unspecified \\
        14 & Moody’s~\cite{moodys_2023} & 2023 & Blockchain & Fiat currencies & Pegged \\
        15 & Decrypt~\cite{decrypt_2023} & 2023 & Unspecified & Fiat currency & Pegged \\
        16 & PwC~\cite{pwc_2022} & 2022 & Unspecified & An asset considered to have a stable value (for instance, a fiat currency or precious metals) & Minimise volatility \\
        17 & KPMG and Aspen Digital~\cite{kpmg_2022} & 2022 & Unspecified & Unspecified & Unspecified \\
        18 & Bluechip~\cite{bluechip_2022} & 2022 & Unspecified & Unspecified & Unspecified \\
        19 & Stellar and Wirex~\cite{stellar_2021} & 2021 & Unspecified & A stable asset & Mitigate the price volatility \\
        20 & Castle Island Ventures~\cite{castle_report} & 2020 & Public blockchain & Sovereign currencies & Track the return of sovereign currencies \\
        \bottomrule
    \end{tabular}
    }
    \captionsetup{position=below}
    \caption{Stablecoin definitions from industry reports. Note that all descriptions in the last three columns are directly \emph{quoted} from the original text, except for the ``Unspecified''s.}
    \label{tab:def_ind}
\end{table*}

\subsubsection{Top Web3 Media}
\label{sec:web3_media}

The top Web3 media list, shown in Table~\ref{tab:web3_media}, is determined by the total visits in the recent 2 years (December 2022 - November 2024), according to Semrush Traffic Analytics~\cite{semrush}.

\subsection{Existing Stablecoins}
\label{app:existing_stablecoins}

The list of 95 active stablecoins with a market capitalization exceeding \$10M (May 2025) is shown in Table~\ref{tab:stablecoins1},~\ref{tab:stablecoins2}.

\begin{table*}[p]
    \scriptsize
    \centering
    \begin{tabular}{lp{2cm}lp{1cm}p{0.7cm}|p{2cm}|p{2.2cm}|p{0.9cm}p{4cm}}
        \toprule
        No. & Project & Stablecoin & Market cap & Pegged asset & Collateral asset & Stabilization mechanism & Yield rate & Yield source \\
        \midrule
        1 & Tether & USDT & \$152,797M & USD & Fiat currency & Implicit & N/A \\
        2 & Circle & USDC & \$61,523M & USD & Fiat currency & Implicit & N/A \\
        3 & Sky & USDS & \$7,007M & USD & Cryptocurrency & Liquidation and Emergency & 6.5\% & Native protocol revenue, cash and cash equivalents yield, and external DeFi protocol yield \\
        4 & Ethena Labs & USDe & \$5,216M & USD & Cryptocurrency & Hedging & 4\% & L1 staking reward, derivatives-driven yield, and third-party custodian revenue\\
        5 & MakerDAO & DAI & \$4,539M & USD & Cryptocurrency & Liquidation & N/A \\
        6 & World Liberty Financial & USD1 & \$2,152M & USD & Fiat currency & Implicit & N/A \\
        7 & First Digital Labs & FDUSD & \$1,628M & USD & Fiat currency & Implicit & N/A \\
        8 & Ethena Labs & USDtb & \$1,443M & USD & Fiat currency & Implicit & N/A \\
        9 & PayPal & PYUSD & \$904M & USD & Fiat currency & Implicit & N/A \\
        10 & Usual & USD0 & \$635M & USD & Fiat currency & Implicit & 10\% & Cash and cash equivalents yield, and secondary token emission \\
        11 & Ondo Finance & USDY & \$580M & USD & Fiat currency & Implicit & 4.35\% & Cash and cash equivalents yield \\
        12 & TrueUSD & TUSD & \$494M & USD & Fiat currency & Implicit & N/A \\
        13 & Maple Finance & SyrupUSDC & \$456M & USD & Cryptocurrency & Liquidation & 10.1\% & Native protocol revenue \\
        14 & Hashnote & USYC & \$390M & USD & Fiat currency & Implicit & Not mentioned & Cash and cash equivalents yield \\
        15 & Falcon Finance & USDf & \$384M & USD & Cryptocurrency & Liquidation, hedging, and supply adjustment & 9.4\% & Derivatives-driven yield and external DeFi protocol yield \\
        16 & USDD & USDD & \$376M & USD & Cryptocurrency & Supply adjustment & 20\% & Community-subsidized fund \\
        17 & Stables Labs & USDX & \$373M & USD & Cryptocurrency & Hedging & 8.23\% & Derivatives-driven yield and community-subsidized fund \\
        18 & Ripple & RLUSD & \$310M & USD & Fiat currency & Implicit & N/A \\
        19 & Global Dollar Network & USDG & \$278M & USD & Fiat currency & Implicit & N/A \\
        20 & Resolv Labs & USR & \$259M & USD & Cryptocurrency & Hedging & 8\% & L1 staking reward and derivatives-driven yield \\
        21 & Aave & GHO & \$250M & USD & Cryptocurrency & Liquidation and supply adjustment & 4.23\% & Community-subsidized fund \\
        22 & Blast & USDB & \$244M & USD & Cryptocurrency & Implicit & 13.5\% & Third-party custodian revenue \\
        23 & M0 & M & \$237M & USD & Fiat currency & Supply adjustment & 4.32\% & Cash and cash equivalents yield \\
        24 & Circle & EURC & \$235M & EUR & Fiat currency & Implicit & N/A \\
        25 & OpenEden & USDO & \$227M & USD & Fiat currency & Implicit & 3.9\% & Cash and cash equivalents yield \\
        26 & Avalon Labs & USDa & \$201M & USD & Cryptocurrency & Liquidation & 5\% & Native protocol revenue \\
        27 & Level & lvlUSD & \$184M & USD & Cryptocurrency & Supply adjustment & 10.72\% & External DeFi protocol yield \\
        28 & Elixir & deUSD & \$184M & USD & Cryptocurrency & Hedging & 5.79\% & Derivatives-driven yield and third-party custodian revenue\\
        29 & Curve Finance & crvUSD & \$168M & USD & Cryptocurrency & Supply adjustment and emergency & 1.1\% & Native protocol revenue \\
        30 & A7A5 & A7A5 & \$143M & RUB & Fiat currency & Implicit & 8.63\% & Cash and cash equivalents yield \\ 
        31 & Stasis & EURS & \$140M & EUR & Fiat currency & Implicit & N/A \\
        32 & Paxos & USDL & \$137M & USD & Fiat currency & Implicit & 3.7\% & Cash and cash equivalents yield \\
        33 & Aster & USDF & \$137M & USD & Cryptocurrency & Supply adjustment and hedging & 5.9\% & Derivatives-driven yield and secondary token emission \\
        34 & Agora & AUSD & \$126M & USD & Fiat currency & Implicit & N/A \\
        35 & Anzen & USDz & \$122M & USD & RWA & Supply adjustment & 15.7\% & Third-party custodian revenue \\
        36 & Resupply & reUSD & \$120M & USD & Cryptocurrency & Supply adjustment and liquidation & 21.77\% & Native protocol revenue and secondary token emission \\
        37 & Berachain & HONEY & \$88M & USD & Cryptocurrency & Implicit & N/A \\
        38 & Inverse Finance & DOLA & \$79M & USD & Cryptocurrency & Supply adjustment & 8.75\% & Native protocol revenue \\
        39 & Reservoir & rUSD & \$75M & USD & Fiat currency, RWA, and cryptocurrency & Supply adjustment & 8.5\% & Community-subsidized fund \\
        40 & Paxos & USDP & \$72M & USD & Fiat currency & Emergency & N/A \\
        41 & Frax Finance & frxUSD & \$68M & USD & Fiat currency & Implicit & 7.15\% & Derivatives-driven yield, external DeFi protocol yield, and cash and cash equivalents yield \\
        42 & Bucket Protocol & BUCK & \$65M & USD & Cryptocurrency & Liquidation and supply adjustment & 27.21\% & Not mentioned \\
        43 & Avant & avUSD & \$62M & USD & Cryptocurrency & Hedging & 7.99\% & Derivatives-driven yield \\
        44 & Cygnus Finance & cgUSD & \$60M & USD & Fiat currency & Implicit & 4.25\% & Cash and cash equivalents yield \\
        45 & Anchored Coins & AEUR & \$58M & EUR & Fiat currency & Implicit & N/A \\
        46 & Binance & BUSD & \$57M & USD & Cryptocurrency & Implicit & N/A \\
        47 & Abracadabra & MIM & \$55M & USD & Cryptocurrency & Supply adjustment & 17.68\% & External DeFi protocol yield \\
        48 & Felix & feUSD & \$51M & USD & Cryptocurrency & Liquidation and supply adjustment & 16.93\% & Native protocol revenue \\
        49 & Lista & lisUSD & \$50M & USD & Cryptocurrency & Liquidation and supply adjustment & 6.86\% & Not mentioned \\
        50 & Web 3 Dollar & USD3 & \$49M & USD & Cryptocurrency & Implicit & 4.4\% & Not mentioned \\
        51 & Gemini & GUSD & \$49M & USD & Fiat currency & Implicit & N/A \\
        52 & Overnight Finance & USD+ & \$48M & USD & Cryptocurrency & Implicit & 16.21\% & Native protocol revenue, derivatives-driven yield, and external DeFi protocol yield\\
        53 & Transfero & BRZ & \$48M & BRL & Fiat currency & Implicit & N/A \\
        54 & Mountain Protocol & USDM & \$47M & USD & Fiat currency & Implicit & 4.5\% & Cash and cash equivalents yield \\
        55 & Banking Circle & EURI & \$47M & EUR & Fiat currency & Implicit & N/A \\
        56 & Societe Generale & EURCV & \$46M & EUR & Fiat currency & Implicit & N/A \\
        \bottomrule
    \end{tabular}
    \captionsetup{position=below}
    \caption{The list of active stablecoins with a market capitalization exceeding \$10M (as of May 2025).}
    \label{tab:stablecoins1}
\end{table*}

\begin{table*}[p]
    \scriptsize
    \centering
    \begin{tabular}{lp{2cm}lp{1cm}p{0.7cm}|p{2cm}|p{2.2cm}|p{0.9cm}p{4cm}}
        \toprule
        No. & Project & Stablecoin & Market cap & Pegged asset & Collateral asset & Stabilization mechanism & Yield rate & Yield source \\
        \midrule
        57 & f(x) & fxUSD & \$46M & USD & Cryptocurrency & Emergency & 12.1\% & L1 staking reward and derivatives-driven yield \\
        58 & Rings & scUSD & \$42M & USD & Cryptocurrency & Implicit & 8.2\% & External DeFi protocol yield \\
        59 & Liquity & LUSD & \$41M & USD & Cryptocurrency & Supply adjustment & N/A \\
        60 & Tether & EURT & \$40M & EUR & Fiat currency & Implicit & N/A \\
        61 & Celo & CUSD & \$35M & USD & Cryptocurrency & Supply adjustment & N/A \\
        62 & Hive & HBD & \$34M & USD & Cryptocurrency & Supply adjustment & 20\% & Community-subsidized fund \\
        63 & StraitsX & XUSD & \$34M & USD & Fiat currency & Implicit & N/A \\
        64 & Plume & pUSD & \$29M & USD & Cryptocurrency & Implicit & N/A \\
        65 & Liquity & BOLD & \$28M & USD & Cryptocurrency & Supply adjustment & 7.94\% & Native protocol revenue \\
        66 & Monerium & EURe & \$27M & EUR & Fiat currency & Implicit & N/A \\
        67 & MNEE & MNEE & \$26M & USD & Fiat currency & Implicit & N/A \\
        68 & Angle & USDA & \$26M & USD & Cryptocurrency & Supply adjustment & 5.53\% & Third-party custodian revenue, native protocol revenue, and external DeFi protocol yield \\
        69 & Hex Trust & USDX & \$25M & USD & Fiat currency & Implicit & N/A \\
        70 & Synthetix & sUSD & \$24M & USD & Cryptocurrency & Implicit & N/A \\
        71 & Gyroscope & GYD & \$24M & USD & Cryptocurrency & Supply adjustment and emergency & 11.56\% & Not mentioned \\  
        72 & BiLira & TRYB & \$24M & TRY & Fiat currency & Implicit & N/A \\
        73 & StandX & DUSD & \$23M & USD & Cryptocurrency & Hedging and emergency & 8.46\% & L1 staking reward and derivatives-driven yield \\
        74 & River & satUSD & \$23M & USD & Cryptocurrency & Liquidation & 20.57\% & protocol fees \\
        75 & Noon & USN & \$22M & USD & Fiat currency and cryptocurrency & Hedging & 7.38\% & Derivatives-driven yield, and cash and cash equivalents yield \\
        76 & Angle & EURA & \$22M & EUR & Cryptocurrency & Supply adjustment & 4.31\% & Cash and cash equivalents yield, external DeFi protocol yield, and native protocol revenue \\
        77 & Electronic Dollar & EUSD & \$21M & USD & Cryptocurrency & Implicit & 4.3\% & Not mentioned \\
        78 & GMO Trust & ZUSD & \$19M & USD & Fiat currency & Implicit & N/A \\
        79 & Aegis & YUSD & \$18M & USD & Cryptocurrency & Hedging and supply adjustment & 11\% & Derivatives-driven yield \\
        80 & Yala & YU & \$16M & USD & Cryptocurrency & Liquidation and supply adjustment & 9.43\% & Native protocol revenue, external DeFi protocol yield, and cash and cash equivalents yield \\
        81 & dForce & USX & \$15M & USD & Cryptocurrency & Emergency & 8\% & Community-subsidized fund \\
        82 & Solayer & sUSD & \$15M & USD & Fiat currency & Implicit & 3.99\% & Cash and cash equivalents yield \\
        83 & PT Rupiah Token & IDRT & \$14M & IDR & Fiat currency & Implicit & N/A \\
        84 & Alchemix & alUSD & \$14M & USD & Cryptocurrency & Implicit & N/A \\
        85 & Frankencoin & ZCHF & \$13M & CHF & Cryptocurrency & Liquidation & 4.175\% & Community-subsidized fund \\
        86 & GMO Trust & GYEN & \$13M & JPY & Fiat currency & Implicit & N/A \\
        87 & Orby & USC & \$12M & USD & Cryptocurrency & Liquidation & 0\% & Native protocol revenue and secondary token emission\\
        88 & StablR & EURR & \$11M & Euro & Fiat currency & Implicit & N/A \\
        89 & Youves & uUSD & \$11M & USD & Cryptocurrency & Liquidation & 14\% & Secondary token emission \\
        90 & Moneta & USDM & \$11M & USD & Fiat currency & Implicit & N/A \\
        91 & WSPN & WUSD & \$11M & USD & Fiat currency & Implicit & N/A \\
        92 & JUST & USDJ & \$10M & USD & Cryptocurrency & Liquidation & N/A \\
        93 & StraitsX & XSGD & \$10M & SGD & Fiat currency & Implicit & N/A \\
        94 & defi.money & MONEY & \$10M & USD & Cryptocurrency & Liquidation & 26.90\% & Native protocol revenue \\
        95 & Anzens & USDA & \$10M & USD & Fiat currency & Implicit & N/A \\
        \bottomrule
    \end{tabular}
    \captionsetup{position=below}
    \caption{(Cont'd) The list of active stablecoins with a market capitalization exceeding \$10M (as of May 2025).}
    \label{tab:stablecoins2}
\end{table*}

\subsection{Security Incidents}
\label{app:security_incidents}

The dataset of stablecoin security incidents with losses exceeding \$100K are detailed in Table~\ref{tab:security_incidents}.

\begin{table*}[!tb]
    \centering
    \scriptsize
    \begin{tabular}{lllllll}
        \toprule
        No. & Project & Stablecoin & Blockchain & Year & Loss & Root cause \\
        \midrule
        1 & Terra & UST & Terra & 2022 & \$40B & Market fluctuation \\
        2 & Neutrino & USDN & Waves & 2022 & \$200M & Market fluctuation \\
        3 & Beanstalk & Bean & Ethereum & 2022 & \$182M & Flash loan attack and governance attack \\
        4 & BonqDAO & BEUR & Polygon & 2023 & \$120M & Price manipulation \\
        5 & Cashio & CASH & Solana & 2022 & \$53M & Code vulnerability \\
        6 & DEFI100 & D100 & BSC & 2021 & \$32M & Rug pull \\
        7 & Mochi & USDM & Ethereum & 2021 & \$30M & Rug pull and governance attack \\
        8 & Tether & USDT & Ethereum & 2017 & \$31M & Access control \\
        9 & UwU Lend & sUSDe & Ethereum & 2024 & \$23M & Price manipulation \\
        10 & Angle Protocol & EURA\&USDA & Ethereum & 2023 & \$18M & Impacted fund \\
        11 & Deus Finance & DEI & Fantom & 2022 & \$13M & Flash loan attack and price manipulation \\
        12 & Prisma Finance & mkUSD & Ethereum & 2024 & \$12M & Code vulnerability \\
        13 & Defrost Finance & H2O & Avalanche & 2022 & \$12M & Flash loan attack, access control, and price manipulation \\
        14 & Elephant Money & TRUNK & BSC & 2022 & \$12M & Flash loan attack \\
        15 & Yearn Finance & yUSDT & Ethereum & 2023 & \$11M & Code vulnerability \\
        16 & Platypus Finance & USP & Avalanche & 2023 & \$8.5M & Price manipulation \\
        17 & Haven Protocol & xUSD & Haven & 2021 & \$8.2M & Code vulnerability  \\
        18 & Origin Protocol & OUSD & Ethereum & 2020 & \$8.0M & Code vulnerability \\
        19 & True Seigniorage Dollar & TSD & BSC & 2021 & \$7.1M & Governance attack \\
        20 & Abracadabra Money & MIM & Ethereum & 2024 & \$6.5M & Code vulnerability\\
        21 & Deus Finance & DEI & BSC, Arbitrum & 2023 & \$6.3M & Code vulnerability \\
        22 & Seneca & senUSD & Ethereum, Arbitrum & 2024 & \$6.0M & Code vulnerability \\
        23 & XSURGE & xUSD & BSC & 2021 & \$5.6M & Flash loan attack \\
        24 & Nirvana & NIRV & Solana & 2022 & \$3.5M & Flash loan attack \\
        25 & Raft & R & Ethereum & 2023 & \$3.3M & Flash loan attack \\
        26 & Deus Finance & DEI & Fantom & 2022 & \$3.0M & Flash loan attack and price manipulation \\ 
        27 & Zunami Protocol & UZD & Ethereum & 2023 & \$2.2M & Flash loan attack and price manipulation \\
        28 & Hope Finance & HOPE & Arbitrum & 2023 & \$1.9M & Rug pull \\
        29 & Acala & aUSD & Polkadot & 2022 & \$1.6M & Code vulnerability \\
        30 & Minterest & mUSDY & Mantle & 2024 & \$1.5M & Flash loan attack and code vulnerability \\
        31 & Hubble Protocol & USDH & Solana & 2022 & \$1.3M & Price manipulation \\
        32 & PalmSwap & USDP & BSC & 2023 & \$900K & Code vulnerability \\
        33 & bDollar & BDO & BSC & 2022 & \$730K & Price manipulation \\
        34 & UPFI Network & UPFI & BSC & 2024 & \$521K & Code vulnerability \\
        35 & Anzen Finance & USDz & Base & 2024 & \$500K & Code vulnerability \\
        36 & PolBase Cash & PBC & Ethereum & 2021 & \$354K & Rug pull \\
        37 & SperaxUSD & USDs & Arbitrum & 2023 & \$300K & Code vulnerability \\
        38 & TheStandard.io & PAXG & Arbitrum & 2023 & \$290K & Price manipulation \\
        39 & Kujira Network & USK & Kujira & 2023 & \$260K & Code vulnerability \\
        40 & Safe Dollar & SDO & Polygon & 2021 & \$248K & Flash loan attack \\
        41 & Linear Finance & LUSD & BSC & 2023 & \$212K & Code vulnerability \\
        42 & Iron Finance & IRON & BSC & 2020 & \$170K & Code vulnerability \\
        43 & Elephant Money & TRUNK & BSC & 2023 & \$165K & Price manipulation \\
        44 & Abracadabra Money & MIM & Ethereum & 2022 & \$111K & Price manipulation \\
        \bottomrule
    \end{tabular}
    \captionsetup{position=below}
    \caption{Existing security incidents of stablecoins with losses exceeding \$100K.}
    \label{tab:security_incidents}
\end{table*}

\end{document}